\pdfoutput=1
\documentclass[twocolumn,aps,prb,footinbib,floatfix,nolongbibliography,superscriptaddress]{revtex4-2}
\usepackage[utf8]{inputenc}

\usepackage{amsmath,amssymb,bm}
\usepackage{xcolor,graphicx,siunitx}
\newcommand{\Tr}{\operatorname{Tr}}
\definecolor{dodgerblue}{HTML}{1E90FF}
\definecolor{columbiablue}{HTML}{87AFC7}
\usepackage[colorlinks=true,citecolor=dodgerblue,linkcolor=dodgerblue,urlcolor=dodgerblue]{hyperref}
\usepackage{tikzsymbols}
\usepackage{slashed}

\begin{document}

\title{\textbf{Vacancy-driven topological transition of a Chern insulator}}
\author{Peize Ding}
\email{pd2714@columbia.edu}
\affiliation{Department of Physics, Columbia University, New York, NY, 10027, USA}
\author{Raquel Queiroz}
\email{raquel.queiroz@columbia.edu}
\affiliation{Department of Physics, Columbia University, New York, NY, 10027, USA}
\affiliation{Center for Computational Quantum Physics, Flatiron Institute, New York, New York 10010, USA}
\date{\today}

\begin{abstract}
We develop a microscopic description of the strong-disorder transition in a Chern insulator using the in-gap ring states bound to individual vacancies. 
Using Green's function formalism, we construct an energy-dependent Hamiltonian describing their hybridization. 
 In a commensurate periodic vacancy lattice, a Dirac inversion of the zero-energy vacancy kernel accounts for the change in the occupied-state Chern number of the full system. 
We treat weak positional disorder using the self-consistent Born approximation and numerical disorder averaging, and study independently distributed vacancies using transfer-matrix and multifractal scaling. 
The local Chern marker and current textures track the transition in real space. Together, these results highlight the role that nonperturbative ring states have as microscopic degrees of freedom through which strong disorder destroys the Chern phase.
\end{abstract}

\maketitle

\paragraph*{Introduction.---}Disorder-driven topological transitions are diagnosed through the emergence of critical delocalized states: most bulk states are localized, but at the transition the localization length diverges and the Hall conductance changes by an integer~\cite{Klitzing1980IQHE,PrangeGirvinBook,TKNN1982,Niu1985,Kohmoto1985,Ando1974Localization,Trugman1983Critical,Huckestein1995Scaling,lee1981anderson}.
This paradigm is well established in the integer quantum Hall effect and extends to Chern insulators, where topology originates from Berry curvature rather than applied magnetic fields~\cite{Qi2006,Onoda2003,haldane1988model,Sheng2006ChernTransition,moreno2023topological,AgarwalaShenoy2017}. These transitions are well described by field theories for weak disorder, network models in the semiclassical regime, and large-scale numerical simulations~\cite{Sondhi1997QHEReview,Pruisken1984NLsigma,ChalkerCoddington1988,Huckestein1995Scaling,mackinnon1983scaling,Evers2008AndersonReview}. Recent developments in band geometry and topology---particularly the understanding that topological bands emerge from real-space orbital interference at unit-cell scales---raise the question of whether there is also a universal real-space understanding of how a topological transition occurs in the presence of strong disorder.

\begin{figure}[tbp]
\centering
\includegraphics[width=1.0\columnwidth]{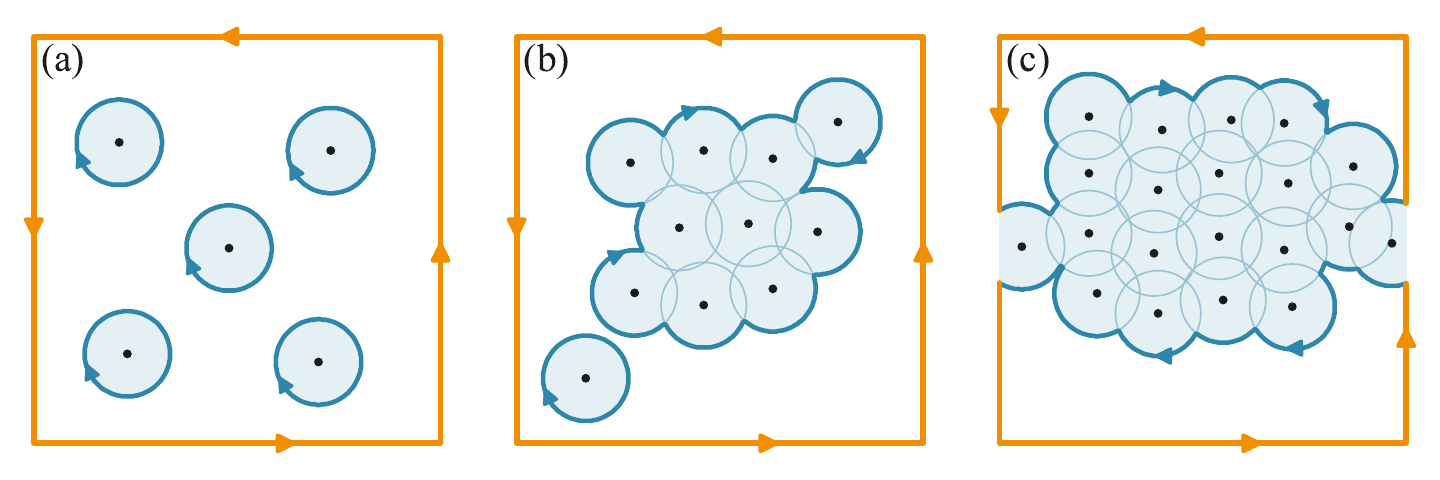}
\caption{
Schematic real-space picture for a vacancy-disordered Chern insulator: black dots mark vacancies, light-blue circles denote the bound ring states, and the orange boundary mode is the parent chiral edge state.
(a) Dilute vacancies support isolated ring states, and the edge mode persists.
(b) Overlapping ring states hybridize into a connected cluster with a chiral boundary mode.
(c) A spanning cluster reconnects the chiral modes into separate loops rather than a mode encircling the full sample.
Hybridization of the ring states produces an impurity band. In the periodic vacancy lattice, its inversion contributes $\Delta C=-1$, opposite to the parent bulk's $C=+1$, leaving a topologically trivial state with $C=0$.
}
\label{fig:percolation_demonstration}
\end{figure}

Here, we provide a microscopic description of the vacancy-driven topological transition in terms of the proliferation and hybridization of ring states~\cite{slager2015impurity,queiroz2024ring,Tornow2026Observation}. 
Although vacancies are not topological defects, strong local potentials nucleate in-gap states due to the topology of the bulk. In any Chern band, these states carry a circulating current determined by the bulk topology of the clean parent Hamiltonian. They are nonperturbative consequences of multiple scattering events at the same impurity site and appear as low-energy poles of the T-matrix that persist to arbitrarily large impurity strengths~\cite{queiroz2024ring}. At finite vacancy density, the ring states hybridize into an impurity band; however, this band behaves unconventionally. The overlap of the ring states suggests a percolation picture for the topological transition (see Fig.~\ref{fig:percolation_demonstration}).

In this Letter, we study the vacancy limit of the topological phase transition by first considering a periodic vacancy lattice, for which the impurity band is translationally invariant. This limit allows us to identify the critical impurity concentration required to close the impurity-band gap and cause a band inversion that cancels the clean bulk topology. The resonant lattice of vacancy states forms a Dirac dispersion carrying a $\pi$ Berry phase. The location of this transition depends only on the microscopic details encoded in the clean resolvent Green's function, $G_0(\epsilon)=(\epsilon-H)^{-1}$, and not on the local potential itself. We then introduce random vacancy displacements, whose effect can be captured perturbatively about the ordered vacancy lattice using the self-consistent Born approximation. Full numerical simulations of independently distributed vacancies establish the strong-disorder transition directly.

\paragraph*{{Low-energy} ring states.---}
We begin with the Qi--Wu--Zhang (QWZ) model~\cite{Qi2006}, a prototypical Chern insulator with Bloch Hamiltonian $h_0(\mathbf{k})=v\sin k_x\sigma_x-v\sin k_y\sigma_y+[m-t(\cos k_x+\cos k_y)]\sigma_z$, where $\boldsymbol{\sigma}$ acts on the $s,p$ orbitals and $t=2$ sets the energy unit. Its Chern number is $C=1$ ($C=-1$) for $0<m/2t<1$ ($-1<m/2t<0$), and zero otherwise~\cite{TKNN1982,Niu1985,Kohmoto1985,Xiao2010BerryReview}.
Writing $c_{\mathbf r}=(c_{\mathbf r,s},c_{\mathbf r,p})^T$, the clean real-space Hamiltonian is
$
\hat H_0=\sum_{\mathbf r,\mathbf r'}c_{\mathbf r}^{\dagger}h_0(\mathbf r-\mathbf r')c_{\mathbf r'},
$
where $h_0(\mathbf r-\mathbf r')$ is its Fourier transform.
An impurity at $\mathbf a_0$ is described by
$
\hat H_{\mathrm{imp}}=c_{\mathbf a_0}^{\dagger}Vc_{\mathbf a_0},
$
with $V=V_0(|s\rangle\langle s|+|p\rangle\langle p|)$, so both orbitals experience the same local potential.
The total Hamiltonian is $\hat H=\hat H_0+\hat H_{\mathrm{imp}}$.

We first ask whether a single defect produces an in-gap state. Its bound-state energy $\epsilon_b$ obeys the $T$-matrix pole condition~\cite{slager2015impurity}
\begin{equation}
\det\!\left[1 - V\, g_0(\epsilon_b; \mathbf{a}_0, \mathbf{a}_0)\right] = 0,
\label{eq:Tmatrix}
\end{equation}
where
\begin{equation}\label{eq:green_real}
g_0(\epsilon; \mathbf{r}, \mathbf{r}') = \langle \mathbf{r} | (\epsilon - h_0 + i0^+)^{-1} | \mathbf{r}' \rangle
\end{equation}
is the clean real-space Green's function, with $g_0(\epsilon;\mathbf r,\mathbf r')=g_0(\epsilon;\mathbf r-\mathbf r')$ by translation symmetry. In the topological phase, a local potential of any strength produces in-gap bound states~\cite{slager2015impurity,queiroz2024ring}. 
Because $V$ acts on both orbitals, there are two such states; their evolution with $1/V_0$ is shown in SI~\ref{sup:bound_state_texture}~\cite{SM}. 
To study the topological transition driven by vacancies, we work in the limit where the vacancies are unitary scatterers $V_0\to\infty$. 
Thus, since $V_0\to\infty$ gives $T_i^{-1}(\epsilon)\to-g_0(\epsilon;0)$, the $T$ matrix of Eq.~\eqref{eq:Tmatrix} has poles in the zeroes of $\det g_0(\epsilon;0)$ for sufficiently far vacancies, i.e. once distinct vacancies' ring states no longer overlap.
Because of the infinite on-site potential, the wavefunction amplitude vanishes at the vacancy center, producing a ring-shaped probability density around $\mathbf a_0$; these states are therefore known as \textit{ring states}~\cite{queiroz2024ring}. 
The parent-band quantum geometry imposes both a vortex-like phase winding and circulating bond currents, as well as a minimal spatial extension, as discussed in  SI~\ref{sup:metric_bound_multiband}~\cite{SM}.

 The QWZ Hamiltonian has exact local particle-hole symmetry (PHS), represented by $\mathcal{C}=\sigma_xK$ with $K$ denoting complex conjugation, which is restored in the strict vacancy limit (SI~\ref{sup:ph_asymmetry}~\cite{SM}).
Consequently, the zero-energy problem belongs to Altland--Zirnbauer class D, whereas each finite-energy sector belongs to class A~\cite{AltlandZirnbauer1997}.

\paragraph*{Impurity-band Hamiltonian.---}
We now consider a finite concentration of vacancies, described by
$
\hat{H}_{\mathrm{imp}} = \sum_{\mathbf a}c_{\mathbf a}^{\dagger}Vc_{\mathbf a},
$
where $\mathbf a$ labels their positions. Each vacancy hosts two in-gap states at $\pm\epsilon_b$. Because the bulk spectrum is gapped, the low-energy problem can be recast entirely in terms of these ring states and their hybridization, defining an emergent impurity band.

To determine the many-vacancy spectrum, we use the multiple-scattering Green's function for pointlike impurities~\cite{avishai1993electron}. In the vacancy limit, the in-gap eigenenergies $\{\epsilon_m\}$ are fixed by the clean Green's function evaluated on the vacancy positions,
\begin{equation}
\det\!\bigl[g_0(\epsilon_m;\mathbf{a}_i,\mathbf{a}_j)\bigr] = 0.
\label{eq:det_condition}
\end{equation}
Thus the parent Green's function entirely determines the impurity band. We show in SI~\ref{sup:many_impurity_green}~\cite{SM} the full multiple-scattering derivation and wavefunction reconstruction.

For $|\epsilon|<\Delta$, we cast the many-vacancy condition as an effective Hamiltonian using the continuum form~\cite{SM}
$
g_0(\epsilon;0)
=
-\frac{\epsilon}{4\tilde{\Delta}}\, I
-\frac{\epsilon_b}{4\tilde{\Delta}_{\epsilon_b}}\, \sigma_z,
$
where $\tilde{\Delta}=\sqrt{\Delta^2-\epsilon^2}$ and $\tilde{\Delta}_{\epsilon_b}=\sqrt{\Delta^2-\epsilon_b^2}$. In the calculations below, however, we use the exact lattice $g_0$. The vacancy condition becomes the self-consistent eigenvalue problem
\begin{equation}\label{eq:eigen_consistent}
\epsilon_m\, c_i^\alpha
=
\sum_{j,\beta}
h_{ij}^{\prime\,\alpha\beta}(\epsilon_m)\, c_j^\beta,
\end{equation}
with
\begin{equation}\label{eq:eff_H}
\begin{split}
    h_{ii}^{\prime\,\alpha\beta}(\epsilon)
&=
\epsilon\,\delta^{\alpha\beta}
+4\tilde{\Delta}\,g_0^{\alpha\beta}(\epsilon;\mathbf 0),
\qquad
\\
h_{ij}^{\prime\,\alpha\beta}(\epsilon)
&=
4\tilde{\Delta}\,
g^{\alpha\beta}_0(\epsilon;\mathbf{a}_i-\mathbf{a}_j),
\quad i\neq j.
\end{split}
\end{equation}
Equations~\eqref{eq:eigen_consistent} and \eqref{eq:eff_H} define the Hamiltonian for the hybridized ring states. For a trial $\epsilon$, we construct $h'(\epsilon)$ and its spectral range $S(\epsilon)$; a physical solution satisfies $\epsilon\in S(\epsilon)$, and its eigenvector reconstructs the in-gap wavefunction as in SI~\ref{sup:many_impurity_green}.

\paragraph*{Topological transition in the impurity band.---}

We first study a periodic vacancy lattice, where the hybridized ring states form an impurity band with well defined momentum. Here, the topological transition is reflected in a band inversion within the impurity band. We arrange vacancies on a square Bravais superlattice of spacing $d$, with concentration $W=d^{-2}$, and Fourier transform Eq.~\eqref{eq:eff_H} in the vacancy-lattice coordinates to obtain
\begin{equation}
    h'(\boldsymbol{\kappa}, d, \epsilon)=\sum_{\mathbf{a}} h'(\mathbf{a},\epsilon)\,e^{-i\boldsymbol{\kappa}\cdot\mathbf{a}},
\end{equation}
where $\mathbf a$ runs over vacancy-superlattice vectors and $\kappa_{x,y}\in[-\pi/d,\pi/d)$ lies in the mini-Brillouin zone.

\begin{figure}[tbp]
\centering
\includegraphics[width=1.0\columnwidth]{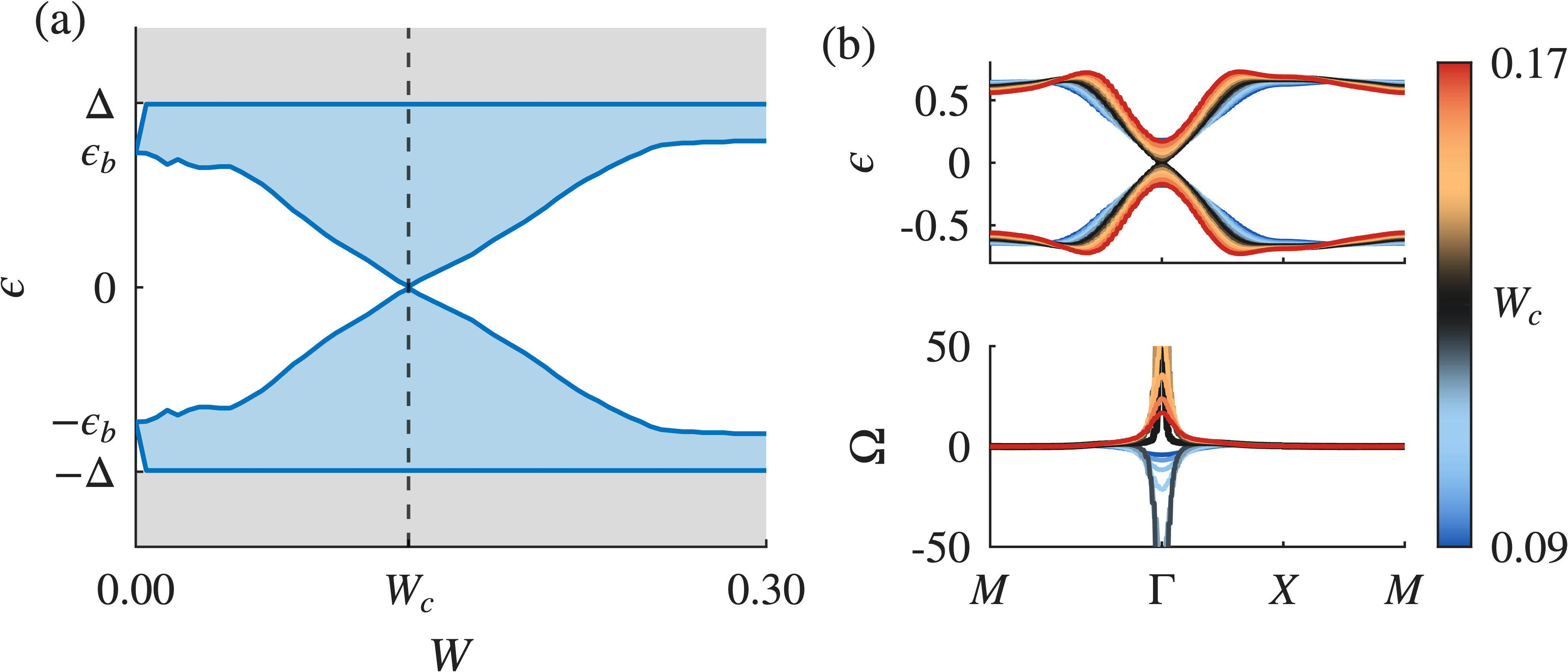}
\caption{
(a) Energy solutions $S_W$ (blue region) as a function of vacancy concentration $W$; the gray region denotes the bulk continuum.
At $W_c \approx 0.120$ the bands of $h'(\epsilon=0)$ touch at $\epsilon=0$, signaling the topological phase transition.
(b) Spectrum and Berry curvature $\Omega$ of $h'(\epsilon=0)$ in the {mini-}Brillouin zone of the ring-state lattice as a function of $W$.
Near $W_c$ (black) the spectrum exhibits a gapped Dirac cone whose mass changes sign, producing a relative Berry-flux transfer of $-2\pi$ at $\Gamma$ in the effective low-energy description.
Parameters: $m=3.5$, $t=2$, $v=0.8$, for which $\Delta\simeq0.462$ and $\epsilon_b\simeq0.337$.}
\label{fig:ring_lattice_percolation}
\end{figure}

Let $S_W$ denote the set of all self-consistent in-gap solutions at concentration $W$.  To follow the in-gap solutions between the discrete vacancy spacings of a lattice-commensurate superlattice, we smoothly interpolate the clean Green's function in distance and evaluate $h'$ at continuous $d$. Figure~\ref{fig:ring_lattice_percolation}(a) shows the resulting evolution. The mini-BZ band structure and representative wavefunctions are shown in SI~\ref{sup:ingap_band_wavefunction}~\cite{SM}. At low $W$ (large $d$), hybridization is weak and the solutions remain near the isolated bound-state energies $\pm\epsilon_b$. As the concentration increases, hybridization strengthens and the impurity bands broaden until they touch at $W_c\approx0.120$ ($d_c\approx2.89$). As shown below, this band touching changes the total Chern number at $\epsilon=0$ from one to zero.

For $h'(0)$, define the on-site, diagonal-hopping, and off-diagonal-hopping amplitudes
$m' \equiv -\epsilon_b\Delta/\tilde{\Delta}_{\epsilon_b}$, $t'(d)=-8\Delta g_0^{ss}(0,d)$, and $v'(d)=8\Delta|g_0^{sp}(0,d)|$, whose $d$ dependence is shown in SI~\ref{sup:Green}~\cite{SM}.
Retaining only nearest-neighbor hopping gives
\begin{equation}\label{eq:Heff_k}
\begin{split}
h'(\boldsymbol{\kappa}, d, \epsilon=0)
=
v'\sin(\kappa_x d)\,\sigma_x
-v'\sin(\kappa_y d)\,\sigma_y\\
+
\bigl[m' - t'\cos(\kappa_x d) - t'\cos(\kappa_y d)\bigr]\sigma_z,
\end{split}
\end{equation}
Up to a global phase rotation, this is a QWZ-type Hamiltonian realized in the emergent impurity band, with parameters promoted to functions of $d$.  Keeping only nearest-neighbor hopping preserves the chirality, but locating the inversion quantitatively requires the longer-range hoppings in $g_0$.
 Its in-plane winding is the same as that of the parent QWZ model, inherited from the phase winding of $g_0^{sp}$ in SI~\ref{sup:Green}. The Dirac inversion changes the Chern number of the negative-eigenvalue band of $h'(0)$ by $-1$ and the total occupied-state Chern number from one to zero (see SI~\ref{sup:periodic_topology_bridge} for the proof for commensurate vacancy supercells).
The impurity band is therefore itself topological: after the inversion, its opposite Chern contribution cancels the topology of the parent bulk.

Including all hoppings, the controlled expansion near $\Gamma$ is
\begin{equation}
h'(\boldsymbol{\kappa}, d, \epsilon=0)
=
	\tilde{v}(d)\,(\kappa_x\sigma_x-\kappa_y\sigma_y)
+
\bigl[\tilde{m}(d) - \tilde{t}(d)\kappa^{2}\bigr]\sigma_z,
\end{equation}
where $\tilde m(d)$, $\tilde v(d)$, and $\tilde t(d)$ are renormalized coefficients incorporating all hopping processes. Figure~\ref{fig:ring_lattice_percolation}(b) shows the evolution of the spectrum and Berry curvature of $h'(0)$ across the transition. At $W_c$, $\tilde m(d)$ and hence the band gap vanish, forming a Dirac cone at $\Gamma$. The Berry curvature is sharply concentrated there and reverses sign as the gap reopens, consistent with the simplified transition criterion in SI~\ref{sup:simplified_transition}~\cite{SM}. This produces the $-2\pi$ relative flux transfer that accounts for the change in the total occupied-state Chern number. The energy-dependent solutions reproduce this local texture where they remain in the gap and merge with the continuum near the mini-BZ boundary, while the complete Berry-curvature and quantum-metric evolution is shown in SI~\ref{sup:ingap_band_wavefunction} and SI~\ref{sup:berry_geometry}~\cite{SM}. Thus increasing vacancy concentration drives an impurity-band inversion, whose real-space counterpart is the connecting defect network.
For the particle-hole asymmetric scalar deformation considered in SI~\ref{sup:ph_asymmetry}~\cite{SM}, the periodic closing shifts away from $\epsilon=0$ but the phenomenology remains the same.

\paragraph*{Topological-Hamiltonian treatment of the disordered transition.---}
We now ask whether it survives when vacancy positions are randomized. Let us write
$
    a_i=\bar a_i+\delta_i ,
$
where $\bar a_i$ is an ideal vacancy-lattice site and $\delta_i$ a random displacement. We denote the root-mean-square displacement of the vacancies from their periodic positions by $\sigma_u$. Positional disorder randomizes the hybridization amplitudes between ring states{, which} becomes a disordered hopping problem within the emergent impurity-band space. Varying $W$ changes the density and connectivity of the emergent ring-state network, rather than simply introducing onsite disorder~\cite{AgarwalaShenoy2017,Sahlberg2020Glassy,BeraDieplingerNayak2024}. The displacement-induced hopping change is
\begin{equation}
\delta h'^{\alpha\beta}_{ij}(\epsilon)
\equiv h'^{\alpha\beta}(\epsilon,a_i-a_j)
-h'^{\alpha\beta}(\epsilon,\bar a_i-\bar a_j),
\end{equation}
where $\alpha,\beta\in\{s,p\}$ and $h'_0(\epsilon)$ is the periodic Hamiltonian.

We next quantify how positional disorder shifts the impurity-band inversion. For physical energy $\epsilon$ entering $h'(\epsilon)$ (the trial energy of Eq.~\eqref{eq:eigen_consistent}, held fixed below) and auxiliary spectral variable $\tilde\epsilon$ (the independent frequency argument of the averaged propagator), we define the disorder-averaged Green's function and topological Hamiltonian
\begin{align}
\overline{g}^{-1}(\tilde\epsilon,\epsilon,\kappa)
&=
\tilde\epsilon+i0^+-h'_0(\epsilon,\kappa)
-\Sigma(\tilde\epsilon,\epsilon,\kappa),
\\
h'_{\rm top}(\epsilon,\kappa)
&=
h'_0(\epsilon,\kappa)
+\mathrm{Re}\,\Sigma(\tilde\epsilon=\epsilon,\epsilon,\kappa).
\end{align}
The topological-Hamiltonian method conjectures that the Chern number of the zero-frequency, disorder-averaged effective Hamiltonian tracks the actual invariant of the disordered system~\cite{wang2010topological,wang2012simplified}, and we apply the same logic here to the disorder-averaged ring-state Hamiltonian. This topological-Hamiltonian construction~\cite{volovik2003universe,gurarie2011single,wang2010topological,wang2012simplified,groth2009theory} restores translation symmetry after averaging. Weak positional disorder renormalizes the impurity-band Dirac mass, therefore shifting the band-inversion point.
Note that we use $h'_{\rm top}$ only as an approximate criterion for this inversion.
Its relation to the sample-resolved Chern invariant and its regime of validity are given in SI~\ref{sup:kernel_SCBA}.

We evaluate $\Sigma$ in two complementary ways. First, we evaluate the positional-disorder kernel exactly to second order in the displacements and solve the resulting self-consistent Born equation (SCBA), with the full kernel given in SI~\ref{sup:kernel_SCBA}~\cite{SM}.
 The periodic vacancy Hamiltonian $h'_0$ treats the vacancy potential nonperturbatively. The Born expansion is in the displacements from the periodic positions. Small displacements change the hybridization between their ring states by $\delta h'\simeq\nabla h'\cdot\boldsymbol\delta_i$. We can therefore expand in displacement without assuming weak vacancy scattering, and the leading disorder-averaged corrections are controlled by $\sigma_u^2$, the mean squared vacancy displacement. For weak positional disorder, the main effect of shifting the vacancies is to change the mass and dispersion of the impurity band.

Second, we obtain the exact disorder-averaged self-energy numerically, retaining the full $\delta h'$. 
For each realization $r$, we construct the full disordered effective Hamiltonian 
$h'_r$ and its Green's function $g_r=[\tilde\epsilon+i0^+-h'_r]^{-1}$. Averaging over realizations gives $\overline g\equiv\langle g_r\rangle_{\rm dis}$, and the Dyson equation then gives $\Sigma_{\rm num}=\tilde\epsilon+i0^+-h'_0-\overline g^{-1}$. 
Using the numerically calculated self-energy, we locate the predicted transition where the Dirac mass vanishes.

\begin{figure}[tbp]
\includegraphics[width=1.0\columnwidth]{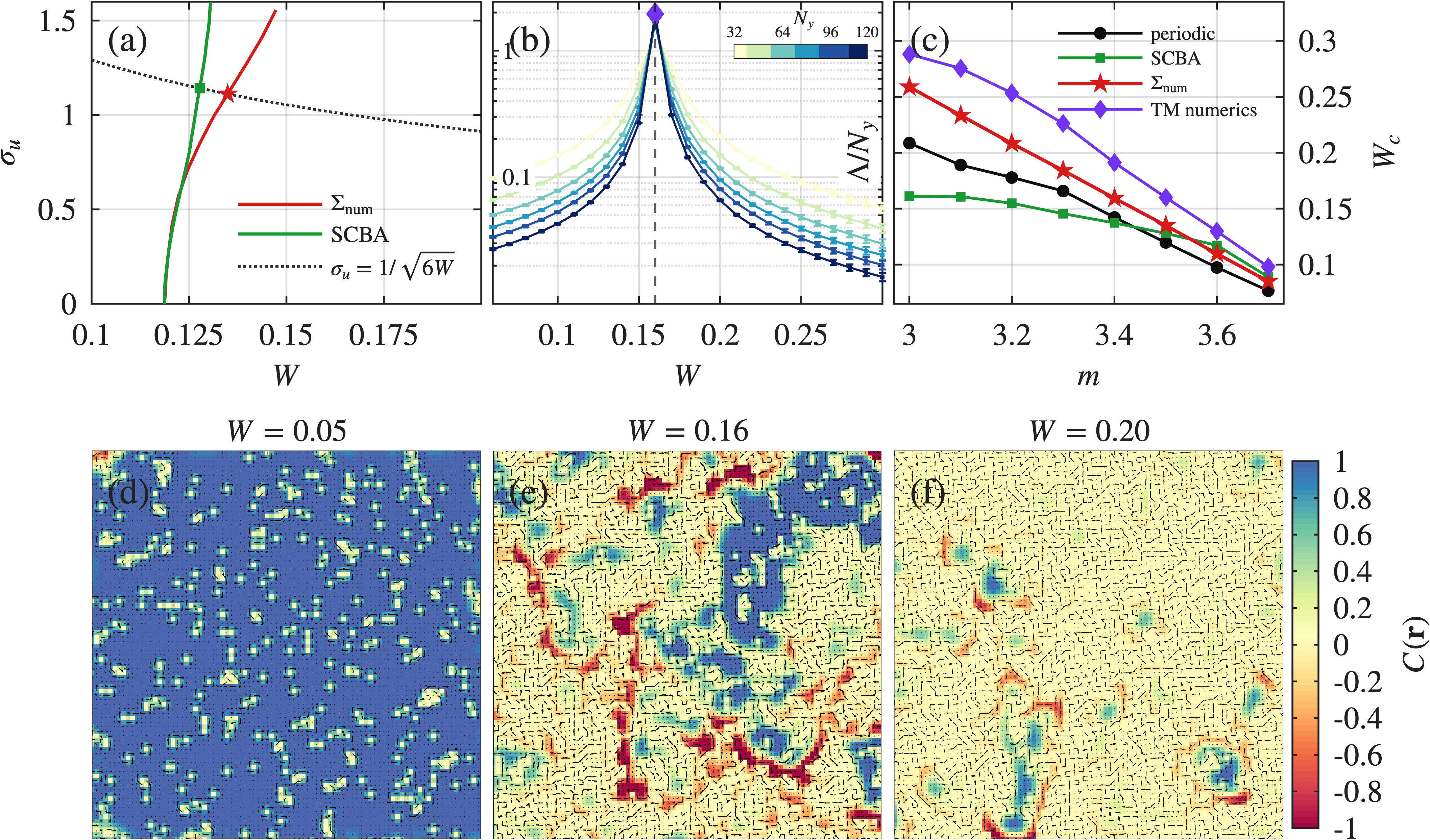}
\caption{
Random defect-band transition ($t=2$, $v=0.8$). (a) Phase boundary in the $(W,\sigma_u)$ plane located by the topological Hamiltonian method for $m=3.5$, with $\sigma_u^2=\langle|\boldsymbol\delta|^2\rangle$. 
Red and green show the numerical disorder-averaged self-energy $\Sigma_{\rm num}$ and SCBA; the dotted line denotes the heuristic density-and-second-moment matching $\sigma_u=1/\sqrt{6W}$ for bimodal vacancies.
(b) Normalized quasi-one-dimensional localization length $\Lambda/N_y$ for the bimodal model, showing a crossing near the critical concentration. 
At $m=3.5$, the critical concentrations obtained from the transfer matrix (TM), the periodic reference, {the SCBA,} and the numerical self-energy are $W_c^{\rm TM}\simeq0.161$, $W_c^{\rm periodic}\simeq0.120$, {$W_c^{\rm SCBA}\simeq0.128$,} and $W_c^{\Sigma_{\rm num}}\simeq0.135$, respectively.
(c) $W_c$ versus $m$, comparing periodic SCBA, SCBA along the matched trajectory, TM, and numerical self-energy. (d)--(f) Local Chern marker (color) and arrow texture at $W=0.05$, $0.16$, and $0.20$, below, near, and above the transition. Definitions are given in SI~\ref{sup:local_chern_current}~\cite{SM}.
}
\label{fig:random_percolation}
\end{figure}

For the parameters studied here, weak positional disorder makes the topological phase survive to higher vacancy concentrations. The two self-energy calculations in Fig.~\ref{fig:random_percolation}(a) differ when the displacements become comparable to the lattice spacing, where higher-order corrections to the Born approximation become important. 

\paragraph*{Numerics on the full disordered Hamiltonian.---}

We now turn from the weakly displaced periodic lattice above to the physically relevant strong-disorder limit, where vacancies are instead placed independently at random. We numerically calculate transfer matrix (TM), local-Chern-marker, and multifractal diagnostics on the original lattice for this bimodal vacancy disorder, defined by independent site potentials $V_i=V_0$ with probability $W$ and $V_i=0$ with probability $1-W$, where large $V_0$ realizes a vacancy.
The TM result is shown in Fig.~\ref{fig:random_percolation}(b), where the scaling of the normalized quasi-one-dimensional localization length $\Lambda/N_y$ shows a crossing at the critical concentration $W_c^{\rm TM}\simeq0.161$ for $m=3.5$ (see SI~\ref{sup:localization_lcm}~\cite{SM} for details).
To compare with the topological Hamiltonian prediction, we map the bimodal ensemble onto an auxiliary periodic vacancy lattice, matching the vacancy density and the second moment of vacancy position within its impurity cell, giving the heuristic trajectory
$
    \sigma_u=\frac{1}{\sqrt{6W}},
$
derived in SI~\ref{sup:localization_lcm}, and its crossing with the phase boundary provides a heuristic comparison of critical concentrations.
Remarkably, the Born approximation, perturbative in the vacancy displacements, already captures the trend in critical vacancy concentration found by transfer-matrix calculations of the full disordered Hamiltonian. Using the numerically calculated self-energy brings the estimate closer to the TM result, giving $W_c^{\Sigma_{\rm num}}\simeq0.135$.
The comparison of the prediction and numerical transition point of the full model across different values of $m$ is shown in Fig.~\ref{fig:random_percolation}(c) (see SI~\ref{sup:kernel_SCBA}~\cite{SM} for the limitations of the topological Hamiltonian approach).

{At $m=3.5$ and $E=0$, we extract a localization-length exponent $\nu\simeq1.3\pm0.1$ from both transfer-matrix scaling and multifractal analysis (SI~\ref{sup:localization_lcm}~\cite{SM}). This value is close to the classical percolation exponent $4/3$~\cite{StaufferAharony1994}, although the accessible system sizes do not establish the asymptotic universality class. Separate fits at $E=0.1$, $0.2$, $0.3$, and $0.4$ give localization-length exponents ranging from $\nu_{\rm TM}=2.18$ to $2.33$ over the accessible system sizes (SI~\ref{sup:finite_energy_exponent}~\cite{SM}). At $E=0$, breaking PHS gives an effective crossing near $W_c\simeq0.139$ with $\nu_{\rm TM}=2.5\pm0.2$ (SI~\ref{sup:finite_energy_exponent}~\cite{SM}).}
{The smaller exponent at zero energy agrees qualitatively with results for particle-hole-symmetric amorphous Chern models~\cite{Sahlberg2020Glassy,ivaki2020criticality}, while breaking this symmetry shifts the exponent toward the class-A integer-quantum-Hall value~\cite{SlevinOhtsuki2009,Sbierski2021DiracIQH,BeraDieplingerNayak2024}. An exact mapping to classical percolation exists for the class-C spin quantum Hall transition~\cite{AltlandZirnbauer1997,GruzbergLudwigRead1999,MaciasGruzbergBettelheim2026}, but the connection for our class-D vacancy problem remains open~\cite{MilnikovSokolov1988,SlevinOhtsuki2009}.}

Fig.~\ref{fig:random_percolation}(d)--(f) show how the topological transition unfolds  in real space. We show both the local Chern marker~\cite{bianco2011mapping,Prodan2017NCG,BellissardElstSchulzBaldes1994} and the arrow texture (SI~\ref{sup:local_chern_current}~\cite{SM}),  which involves exact diagonalization without scaling analysis. In panel (d) the well separated ring states encircle regions of suppressed local Chern marker  which connect until at a critical concentration $W_c$, seen in panel (e),  the region with positive local marker  (blue) no longer percolates through the entire sample. We find that at the transition the local Chern marker fluctuates strongly, and regions  of opposite local-marker sign  appear as well (red regions).  At larger disorder strength, panel (f) most of the sample is trivialized apart from  some isolated islands. 

\paragraph*{Discussion and outlook.}
{We formulate the vacancy-driven topological transition as an impurity-band inversion driven by the hybridization of ring states. In the controlled periodic limit, the full clean-lattice Green's function determines their energies, wavefunctions, and hybridization. The inversion point follows from the zero-energy Green's function evaluated at the vacancy sites. Beyond this inversion, the occupied impurity band contributes a Chern number opposite to that of the parent bulk, changing the total Chern number from $C=1$ to $C=0$. The impurity band is therefore itself topological and compensates the topology of the parent insulator. Even before disorder averaging, the clean Green's function lets us estimate the vacancy concentration where the topological phase is lost. The ring states must hybridize strongly enough for their band to invert. }

{The periodic vacancy lattice treats the strong scattering exactly. We can then expand in small displacements of the vacancies, which change the hybridization between ring states. The Born approximation already captures the trend of the transition found in full-lattice transfer-matrix calculations with independent random vacancies. In real space, the randomly placed ring states hybridize into an irregular network, while regions of suppressed local Chern marker join and the connected region of positive marker breaks apart. This gives a percolation picture of the random transition whose microscopic origin lies in the hybridization of the vacancy-bound states.}

\paragraph*{Acknowledgments.} 
We thank Jia-Xin Zhang{{} and} Daniel Muñoz-Segovia for stimulating discussions.
We acknowledge support from the NSF CAREER award DMR-2340394 and the Alfred P. Sloan Foundation through a Sloan Research Fellowship.

\bibliography{ref_new}

\onecolumngrid
\clearpage
\makeatletter

\begin{center}
\textbf{\large Supplemental Material for ``\@title ''} \\[10pt]

\end{center}
\vspace{20pt}

\setcounter{figure}{0}
\setcounter{section}{0}
\renewcommand{\thesection}{\Roman{section}}
\setcounter{equation}{0}

\renewcommand{\thefigure}{S\@arabic\c@figure}
\renewcommand{\theequation}{S\@arabic\c@equation}
\makeatother

\section{Green's function calculation}\label{sup:Green}

\begin{figure}[htbp]
\centering
\includegraphics[width=1\columnwidth]{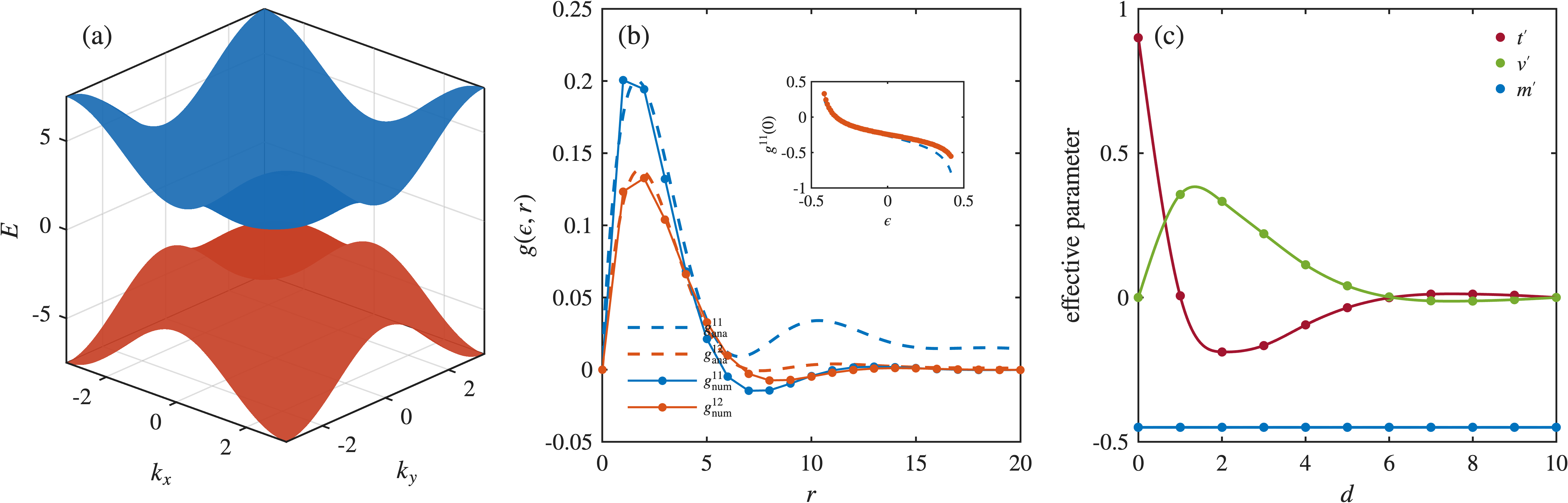}
\caption{(a) Clean QWZ bands in the first Brillouin zone.
(b) The analytical Green's function (Eq.~\eqref{eq:G11}, \eqref{eq:G12}) (dashed line) as a function of $r$ along the $x$-direction, in comparison with the numerical values (dots connected by solid lines).
Inset shows $g_0^{11}(\epsilon, \mathbf{r} = 0)$ as a function of $\epsilon$ using the same dashed/dot-line convention.
For the plotted analytical curves, the finite-cutoff correction in $g_0^{11}$ is fitted as $-A_c e^{-k_c r}$ with $A_c=0.2586$ and $k_c=0.7825$.
For $g_0^{12}$ we keep $k_F=\sqrt{4-m}$ fixed and fit the effective prefactor and damping in $F_1[k_F r(1+i\gamma)]$, giving $C_{12}=0.4869$ and $\gamma=0.5676$.
(c) Distance dependence of the hopping parameters $t', v', m'$ of the effective Hamiltonian at $\epsilon = 0$ {(Eq.~\eqref{eq:eff_para_SI})} extracted from the Green's function.
We use $m = 3.5, t = 2, v = 0.8$.
}
\label{fig:band_Green}
\end{figure}

We introduce a fitted parametrization of the real-space Green's function of the QWZ Hamiltonian. Recall that $h_0(\mathbf{k})=v\sin k_x\sigma_x-v\sin k_y\sigma_y+[m-t(\cos k_x+\cos k_y)]\sigma_z$.
To obtain a compact analytical expression, we use a Fermi-ring parametrization of the Chern-insulator Hamiltonian.
For simplicity, we set $t = 2$ unless otherwise stated.
Then,
\begin{equation}
    h(k,\phi)  =
    \begin{pmatrix} {k}^2 - \mu_\Gamma & \Delta e^{+i\phi} \\ \Delta e^{-i\phi} &  -{k}^2 + \mu_\Gamma\end{pmatrix},
\end{equation}
where the matrix is in the orbital space spanned by $s$ and $p$; we have set $\hbar = 1$, $\phi$ is the azimuthal angle of $\mathbf{k}$, $\Delta$ is an effective gap scale, and $\mu_\Gamma = 4 - m$ sets the nominal ring radius.
{The phase winding of the off-diagonal term is inherited from the QWZ lattice Hamiltonian, while its radial factor is replaced by the effective constant $\Delta$ on the ring. Thus, this is a fitted Fermi-ring parametrization, not a controlled expansion about $\Gamma$; in particular, $\Delta/\mu_\Gamma$ is not small and the exact lattice gap minimum lies away from $\Gamma$. We use it only to describe the spatial form in Fig.~\ref{fig:band_Green}(b), whose analytical curves are fitted directly to the exact lattice $g_0$. All quantitative couplings, transition points, and production results are evaluated using the exact lattice $g_0$.}

The clean Green's function is defined by
\begin{equation}
\begin{split}
        g_0(\epsilon, \mathbf{k}) = \frac{1}{\epsilon - h + i 0^+}
        = \frac{\epsilon\sigma_0 + \xi_\mathbf{k}\sigma_z + \Delta e^{-i\phi} \sigma_- + \Delta e^{i\phi}\sigma_+}{\epsilon^2 - \xi_{\mathbf{k}}^2 - \Delta^2}
\end{split}
\end{equation}
where $\xi_{k} = k^2 - \mu_\Gamma$ { and $\sigma_\pm=(\sigma_x\pm i\sigma_y)/2$, so the upper-right entry of the numerator is $\Delta e^{+i\phi}$, consistent with $h(k,\phi)$ above; the numerator is simply $\epsilon\sigma_0+h$, since $h^2=(\xi_k^2+\Delta^2)\sigma_0$}.
In real space, we first calculate
\begin{equation}
\begin{split}
    g_0^{11}(\epsilon, \mathbf{r}) & = \int\frac{d\mathbf{k}}{(2\pi)^2}
   g_0^{11}(\epsilon, \mathbf{k}) e^{i\mathbf{k}\mathbf{r}}\\
   &= -\frac{1}{8\pi^2}\int d\theta e^{i \sqrt{\mu_\Gamma + \xi} r \cos \theta} \int_{-\mu_\Gamma}^{\Lambda_{\rm UV}} d\xi \frac{\epsilon + \xi}{\xi^2 + \tilde{\Delta}^2}\\
   &= -\frac{1}{8\pi^2}\int d\theta e^{i \sqrt{\mu_\Gamma + \xi} r \cos \theta}(\int_{-\infty}^{\infty} - \int_{-\infty}^{-\mu_\Gamma} - \int_{\Lambda_{\rm UV}}^{\infty}) d\xi \frac{\epsilon + \xi}{\xi^2 + \tilde{\Delta}^2}\\
\end{split}
\end{equation}
where $\Lambda_{\rm UV}$ denotes the cutoff.
The first term
\begin{equation}
    \begin{split}
        FT &= -\frac{1}{8\pi^2}\int_0^{2\pi} d\phi e^{i \sqrt{\mu_\Gamma + \xi} r \cos \phi} \int_{-\infty}^{\infty} d\xi \frac{\epsilon + \xi}{\xi^2 + \tilde{\Delta}^2}\\
        & = -\frac{1}{8\pi^2}\left[\int_0^{\pi} d\phi e^{i \sqrt{\mu_\Gamma + \xi} r \sin \phi} \int_{-\infty}^{\infty} d\xi \frac{\epsilon + \xi}{\xi^2 + \tilde{\Delta}^2} + c.c\right] \\
        & = -\frac{1}{2}\operatorname{Re}\left\{i \operatorname{Res\left[\frac{\epsilon + \xi}{\xi^2 + \tilde{\Delta}^2}\right]_{\xi = i\tilde{\Delta}}} F_0(z)\right\}\\
        &=  -\frac{1}{4}\operatorname{Re}\left\{\left(\frac{\epsilon}{\tilde{\Delta}} + i\right) F_0(z)\right\}
    \end{split}
\end{equation}
where $\tilde{\Delta} \equiv \sqrt{\Delta^2 - \epsilon^2}$ (we work with the energy range $|\epsilon| < \Delta$), $z=k_F r\left(1 + \frac{i\tilde{\Delta}}{2\mu_\Gamma}\right)$, $k_F = \sqrt{\mu_\Gamma}$.
We have used the integral relation
\begin{align}
    \frac{1}{\pi}\int_0^{\pi} d\phi e^{i z \sin \phi} = F_0(z) \equiv J_0(z) + iH_0(z)\\
    \frac{1}{\pi}\int_0^{\pi} d\phi e^{i z \sin \phi + i\phi} = - F_1(z),\qquad
    F_1(z) \equiv J_1(z) - i\left[\frac{2}{\pi}-H_1(z)\right]\label{eq:F1}
\end{align}
Using the fact that $\mu_\Gamma$ and $\Lambda_{\rm UV}$ are large compared to the gap, we evaluate the leftover terms
\begin{equation}
\begin{split}
    & -\frac{1}{8\pi^2}\int d\phi e^{i \sqrt{\mu_\Gamma + \xi} r \cos \phi}(- \int_{-\infty}^{-\mu_\Gamma} - \int_{\Lambda_{\rm UV}}^{\infty}) d\xi \frac{\epsilon + \xi}{\xi^2 + \tilde{\Delta}^2} \\
    &\approx -\frac{1}{8\pi^2}\int d\phi e^{i \sqrt{\mu_\Gamma + \xi} r \cos \phi}(- \int_{-\infty}^{-\mu_\Gamma} - \int_{\Lambda_{\rm UV}}^{\infty}) d\xi \frac{\xi}{\xi^2 + \tilde{\Delta}^2}\\
    &\approx -\frac{1}{4\pi}e^{-k_cr}\int_{-\mu_\Gamma}^{\Lambda_{\rm UV}}d\xi \frac{\xi}{\xi^2 + \tilde{\Delta}^2}\\
    &\approx -\frac{1}{4\pi}e^{-k_cr} \log \frac{\Lambda_{\rm UV}}{\mu_\Gamma}
\end{split}
\end{equation}
In the second line, we used the fact that $\epsilon \ll \xi$ in the integration range.
In the third line, we assumed that these `boundary terms' are only important near $r = 0$ and quickly decay as $r$ becomes large so that we apply an exponentially decaying profile $e^{-k_cr}$ on top of it and get rid of the angular integral.
We also used the fact that $\int_{-\infty}^{\infty} d\xi \frac{\xi}{\xi^2 + \tilde{\Delta}^2} = 0$ because of symmetry.
This gives the leading continuum form, supplemented by a short-distance finite-cutoff correction,
\begin{equation}
    g_0^{11}(\epsilon, \mathbf{r}) = -\frac{1}{4}\operatorname{Re}\left\{\left(\frac{\epsilon}{\tilde{\Delta}} + i\right) F_0(z)\right\}-A_c e^{-k_cr}.
\end{equation}
The constants $A_c$ and $k_c$ parametrize the lattice-scale cutoff dependence of the continuum approximation and are fixed by comparison to the lattice Green's function in Fig.~\ref{fig:band_Green}(b).
{ These constants enter only the analytic continuum form and are not tuned to the transition; the impurity-band inversion itself carries no adjustable parameters.}

Similarly we can look at the off-diagonal term
\begin{equation}
\begin{split}
    g_0^{12}(\epsilon, \mathbf{r}) & = \int\frac{d\mathbf{k}}{(2\pi)^2}
   g_0^{12}(\epsilon, \mathbf{k}) e^{i\mathbf{k}\mathbf{r}}\\
   &= -\frac{1}{8\pi^2}e^{i\theta}\int d\phi e^{i (\sqrt{\mu_\Gamma + \xi} r \cos \phi + \phi)} \int_{-\mu_\Gamma}^{\Lambda_{\rm UV}} d\xi \frac{\Delta}{\xi^2 + \tilde{\Delta}^2}\\
   &= -\frac{1}{8\pi^2}e^{i\theta}\int d\phi e^{i \sqrt{\mu_\Gamma + \xi} r \cos \phi}{e^{i\phi}} (\int_{-\infty}^{\infty} - \int_{-\infty}^{-\mu_\Gamma} - \int_{\Lambda_{\rm UV}}^{\infty}) d\xi \frac{\Delta}{\xi^2 + \tilde{\Delta}^2}\\
   & = -\frac{i}{4}\frac{\Delta}{\tilde{\Delta}}\operatorname{Re}\left\{F_1(z)\right\}e^{i\theta}
\end{split}
\end{equation}
where $\theta$ denotes the azimuthal angle of $\mathbf{r}$ and we have used the definition Eq.~\eqref{eq:F1}.
Here we discarded the boundary term because $g_0^{12}(\epsilon, 0) = 0$ as expected.
$g_0^{21}(\epsilon, \mathbf{r})$ and $g_0^{22}(\epsilon, \mathbf{r})$ can be determined by symmetry.
Finally,
\begin{align}
    g_0^{11}(\epsilon, \mathbf{r}) &= -\frac{1}{4}\operatorname{Re}\left\{\left(\frac{\epsilon}{\tilde{\Delta}} + i\right) F_0(z)\right\} - A_c e^{-k_c r}\label{eq:G11}\\
    g_0^{12}(\epsilon, \mathbf{r}) &= -\frac{i}{4}\frac{\Delta}{\tilde{\Delta}}\operatorname{Re}\left\{F_1(z)\right\}e^{i\theta}\label{eq:G12}\\
    g_0^{21}(\epsilon, \mathbf{r}) &= -g_0^{12*}(\epsilon, \mathbf{r})\\
     g_0^{22}(\epsilon, \mathbf{r}) &= -g_0^{11}(-\epsilon, \mathbf{r})\\
\end{align}
We compare the analytical and numerical Green's function in Fig.~\ref{fig:band_Green} (b), which exhibits very good agreement.

The bound state wavefunction is given by
\begin{equation}
\begin{split}
    \psi^s_{-\epsilon_b}(\mathbf{r}) &= \mathcal{N}
    \begin{pmatrix} g_0^{11}(-\epsilon_b, \mathbf{r}) \\ g_0^{21}(-\epsilon_b, \mathbf{r})\end{pmatrix},
    \end{split}
\end{equation}
\begin{equation}
\begin{split}
    \psi^p_{\epsilon_b}(\mathbf{r}) &= \mathcal{N}
    \begin{pmatrix} g_0^{12}(\epsilon_b, \mathbf{r}) \\ g_0^{22}(\epsilon_b, \mathbf{r})\end{pmatrix},
\end{split}
\end{equation}
The normalization factor can be fixed by
\begin{equation}
    1 = \mathcal{N}^2 \int\frac{d\mathbf{k}}{(2\pi)^2}\frac{(\xi-\epsilon_b)^2 + \Delta^2}{(\xi^2 + \tilde{\Delta}^2)^2}
\end{equation}
For this integral we assume the `boundary term' is not important because the integrand decays fast enough $(\sim \frac{1}{\xi^2})$ so that we can directly implement the residue theorem.
We obtain $\mathcal{N} = \sqrt{\frac{4\tilde{\Delta}^3}{\Delta^2}}$.

{The two in-gap bound states are generated by orbital-polarized source states at the vacancy site,}
\begin{equation}
|e_s\rangle \equiv
\begin{pmatrix} 1 \\ 0 \end{pmatrix},
\qquad
|e_p\rangle \equiv
\begin{pmatrix} 0 \\ 1 \end{pmatrix},
\label{eq:orbitalstructure}
\end{equation}
{in terms of which} the spatial profile can be written compactly as
\begin{equation}
|\psi^\alpha_{\mathbf{a}_0}(\mathbf{r})\rangle = \mathcal{N} g_0(\epsilon_\alpha; \mathbf{r}, \mathbf{a}_0)\, |e_\alpha\rangle,
\label{eq:wf}
\end{equation}
with $\epsilon_\alpha = \mp \epsilon_b$ for $\alpha = s, p$.
{The bound state itself vanishes at the vacancy center, $|\psi^\alpha_{\mathbf{a}_0}(\mathbf{r}=\mathbf{a}_0)\rangle=0$: the off-diagonal component vanishes there because of its phase winding, $g_0^{ps}(\epsilon,\mathbf{0})=0$, while the diagonal component vanishes because the $V_0\to\infty$ bound-state condition is precisely $g_0^{ss}(-\epsilon_b,\mathbf{0})=0$ [equivalently $g_0^{pp}(\epsilon_b,\mathbf{0})=0$], producing the ring-shaped density. Thus $|e_\alpha\rangle$ is the source orbital on which the clean Green's function acts, not the value of the wavefunction at the origin.}
From Eqs.~\eqref{eq:G11}--\eqref{eq:G12} we can summarize the angular dependence as
\begin{align}\label{eq:green_dependence}
    g_0^{ss}(\epsilon, \mathbf{r}) = f_1(\epsilon, r), \quad
     g_0^{sp}(\epsilon, \mathbf{r}) = f_2(\epsilon, r)e^{i\theta},
\end{align}
and $g_0^{pp}(\epsilon, \mathbf{r}) = -g_0^{ss}(-\epsilon, \mathbf{r})$, $g_0^{ps}(\epsilon, \mathbf{r}) = -g_0^{sp*}(\epsilon, \mathbf{r})$, where $f_1$ is real and $f_2$ is purely imaginary.

We also plot the distance dependence of the effective hopping parameters at $\epsilon = 0${,
\begin{equation}\label{eq:eff_para_SI}
    m' \equiv -\epsilon_b\,\frac{\Delta}{\tilde{\Delta}_{\epsilon_b}},\quad
    t'(d) = -8\Delta\, g_0^{ss}(0,d),\quad
    v'(d) = 8\Delta\, |g_0^{sp}(0,d)|,
\end{equation}}
in Fig.~\ref{fig:band_Green} (c).
In this orbital basis, the nearest-neighbor sum gives $h_{\rm NN}^{\prime\,sp}=-8i\Delta f_2(0,d)[\sin(\kappa_xd)+i\sin(\kappa_yd)]$. Since $-i f_2$ is real, its overall sign is absorbed by a momentum-independent relative orbital phase rotation, giving the positive off-diagonal amplitude $v'(d)$ defined above.

\section{Many-vacancy Green's function and wavefunctions}\label{sup:many_impurity_green}

Here we collect the multiple-scattering formulas used to obtain the effective Hamiltonian in the main text.
For a set of pointlike impurities, the exact Green's function can be written as~\cite{avishai1993electron}
\begin{equation}\label{eq:many_imp_Green}
\begin{split}
g(\epsilon;\mathbf{r},&\mathbf{r}')
= g_0(\epsilon;\mathbf{r},\mathbf{r}')+ \sum_{ij} g_0(\epsilon;\mathbf{r},\mathbf{a}_i)\,
\Lambda^{-1}_{ij}(\epsilon)\,
g_0(\epsilon;\mathbf{a}_j,\mathbf{r}'),
\end{split}
\end{equation}
where $i,j$ label impurity positions and
\begin{equation}
\Lambda_{ij}(\epsilon)
=
\frac{1}{T_i(\epsilon)}\delta_{ij}
-\bigl(1-\delta_{ij}\bigr)\,
g_0(\epsilon;\mathbf{a}_i,\mathbf{a}_j).
\end{equation}
The single-impurity $T$ matrix is
$
T_i(\epsilon)=V_i\!\left[I-g_0(\epsilon;0)V_i\right]^{-1}.
$
Thus all multiple scattering between vacancies is encoded in the matrix inverse of $\Lambda$.
Aside from the continuum spectrum of the clean system, the in-gap poles occur when
$
\det \Lambda(\epsilon_m)=0.
$
In the vacancy limit, $V\rightarrow\infty$, one has $T_i^{-1}(\epsilon)\rightarrow -g_0(\epsilon;0)$, so the diagonal and off-diagonal blocks combine into the compact condition
\begin{equation}
\det\!\bigl[g_0(\epsilon_m;\mathbf{a}_i,\mathbf{a}_j)\bigr]=0,
\end{equation}
which is the form used in the main text.

The corresponding in-gap eigenfunctions follow from the pole residue of Eq.~\eqref{eq:many_imp_Green}.
They can be written as a superposition of clean Green's functions sourced at the vacancy positions,
\begin{equation}\label{eq:ingap_wavefunction}
\psi^\alpha(\mathbf{r};\epsilon_m)
=
\sum_{i,\beta}
c^\beta_i\,
g^{\alpha\beta}_0(\epsilon_m;\mathbf{r},\mathbf{a}_i),
\end{equation}
where $c_i^\beta$ is the null vector of $\Lambda(\epsilon_m)$ and $\alpha,\beta\in\{s,p\}$ label the orbital components.
Using the local Green's-function identity in the main text then converts the pole condition into the self-consistent eigenvalue problem Eq.~\eqref{eq:eigen_consistent}, with matrix elements Eq.~\eqref{eq:eff_H}.
For a periodic vacancy lattice, the same equation is Fourier transformed over vacancy-lattice sites to obtain $h'(\boldsymbol{\kappa},d,\epsilon)$.

\subsection{Occupied-state topology for a commensurate vacancy supercell}\label{sup:periodic_topology_bridge}
Since $h'(0)$ is an auxiliary kernel whose eigenvectors are not physical Bloch states, we show here that its band Chern number nonetheless fixes the occupied-state Chern number of the physical vacancy lattice.
For integer $d$ (vacancies on host-lattice sites), the system is periodic with a $d\times d$ supercell. At each supercell momentum $\boldsymbol\kappa$, partition the clean Hermitian Bloch Hamiltonian into retained orbitals and the removed ($s,p$) orbitals of the vacancy site,
\begin{equation}
H=\begin{pmatrix}A&B\\B^\dagger&D\end{pmatrix},\qquad S=D-B^\dagger A^{-1}B.
\end{equation}
Here $A$ is the strict site-removal Hamiltonian, $B$ couples retained and removed orbitals, $D$ is the removed-orbital block, and $S$ is the Schur complement of $A$, i.e., the effective $2\times2$ Hamiltonian of the removed orbitals after the retained ones are integrated out at zero energy [$S^{-1}=(H^{-1})_{vv}$]. $C_{\rm occ}(M)$ denotes the Chern number of the negative-eigenvalue subspace of $M(\boldsymbol\kappa)$ over the mini-BZ, i.e., of the states below $\epsilon_F=0$ for a Hamiltonian; $C_{\rm occ}(H)=1$ for the parent QWZ model. Assume that $H$ and $A$ are smooth and spectrally gapped at zero throughout the mini-BZ, away from a transition. To compare $H$ with the decoupled matrix $\operatorname{diag}(A,S)$, we introduce the deformation parameter $t$ and the family of Hermitian matrices
\begin{equation}
H_t=L_t\begin{pmatrix}A&0\\0&S\end{pmatrix}L_t^\dagger,
\qquad L_t=\begin{pmatrix}I&0\\tB^\dagger A^{-1}&I\end{pmatrix},\quad 0\le t\le1,
\end{equation}
which equals $\operatorname{diag}(A,S)$ at $t=0$ (retained and removed orbitals decoupled) and $H$ at $t=1$; explicitly $H_t=\begin{pmatrix}A&tB\\tB^\dagger&S+t^2B^\dagger A^{-1}B\end{pmatrix}$, so $t$ switches on the coupling $B$. This is a smooth gapped interpolation: $L_t$ is nonsingular, so no zero eigenvalue can appear. Thus $C_{\rm occ}(H)=C_{\rm occ}(A)+C_{\rm occ}(S)$. The removed-orbital block of the clean zero-energy resolvent $G(0)=-H^{-1}$ is $G_{vv}(0)=-S^{-1}$, whose negative-energy subspace is the positive-energy subspace of $S$. The full removed-orbital Bloch bundle is trivial, so $C_{\rm occ}(G_{vv})=-C_{\rm occ}(S)$ and
\begin{equation}\label{eq:vacancy_chern_additivity}
C_{\rm occ}(A)=C_{\rm occ}(H)+C_{\rm occ}(G_{vv}(0)).
\end{equation}
Since $h'(0)=4\Delta G_{vv}(0)$ with $\Delta>0$, this gives the required relation to the zero-energy vacancy kernel. It does not require an isolated physical impurity band over the entire mini-BZ, and it does not equate auxiliary and physical Berry curvatures pointwise. Continuous noncommensurate interpolation and disorder averaging are separate approximations to which this proof does not directly apply.

\section{Vacancy bound states and wavefunction texture}\label{sup:bound_state_texture}

\begin{figure}[tbp]
\centering
\includegraphics[width=1.0\columnwidth]{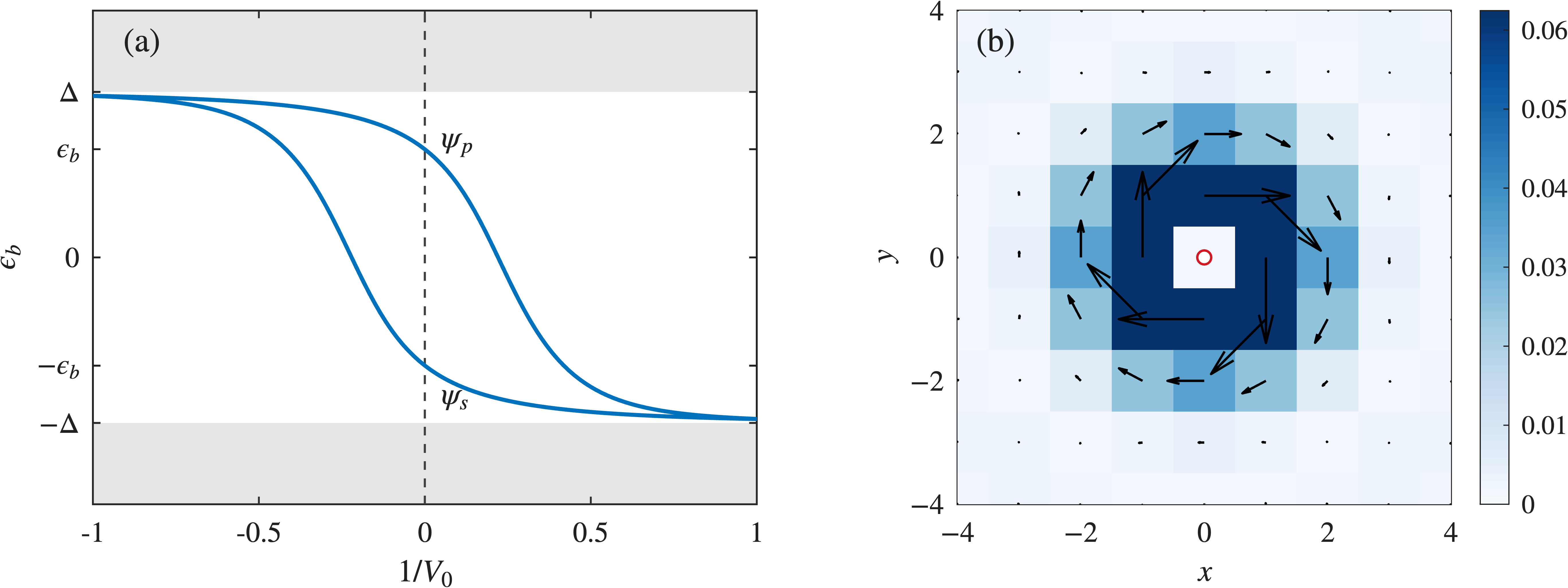}
\caption{(a) Evolution of single-vacancy bound-state energies $\epsilon_b$ as a function of $1 / V_0$.
$\Delta$ denotes the bulk gap, and the shaded region marks the continuum of bulk states.
The two in-gap branches correspond to states of predominantly $s$ and $p$ orbital character, respectively.
(b) Real-space texture of the single-vacancy ring state $\psi^s_{-\epsilon_b}(\mathbf{r})$.
Color denotes $|\psi(\mathbf{r})|^2$, arrows denote the plotted phase-gradient texture $\operatorname{Im}[\psi^\dagger\nabla\psi]$, and the red circle marks the vacancy center where the wavefunction vanishes.
Parameters: $m = 3.5$, $t = 2$, $v = 0.8$.}
\label{fig:bound_state}
\end{figure}

For reference, we collect the single-vacancy bound-state spectrum and wavefunction texture used in the main text in Fig.~\ref{fig:bound_state}.
As shown in Fig.~\ref{fig:bound_state}(a), a local scalar defect in the topological QWZ regime generates two in-gap branches that approach the PHS-related energies $\pm\epsilon_b$ in the strict site-removal limit~\cite{slager2015impurity,queiroz2024ring}. The two branches have predominantly $s$- and $p$-orbital character. Exact spectral PHS holds for the clean and strict site-removal Hamiltonians; a finite scalar potential $V_0\mathbb I$ breaks it, so finite-$V_0$ branches need not be exactly symmetric.

Because of the infinite on-site potential, the wavefunction amplitude vanishes at the vacancy center, producing a ring-shaped probability density surrounding $\mathbf{a}_0$ (\textit{ring state}) \cite{queiroz2024ring}.
The ring-state wavefunction exhibits vortex-like phase winding inherited from the parent topology; physical circulation is carried by the QWZ lattice bond current.
The winding of $\psi^{p}_{\epsilon_b}$ has opposite chirality.

\section{In gap band structure and wavefunction}\label{sup:ingap_band_wavefunction}

\begin{figure}[tbp]
\centering
\includegraphics[width=1.0\columnwidth]{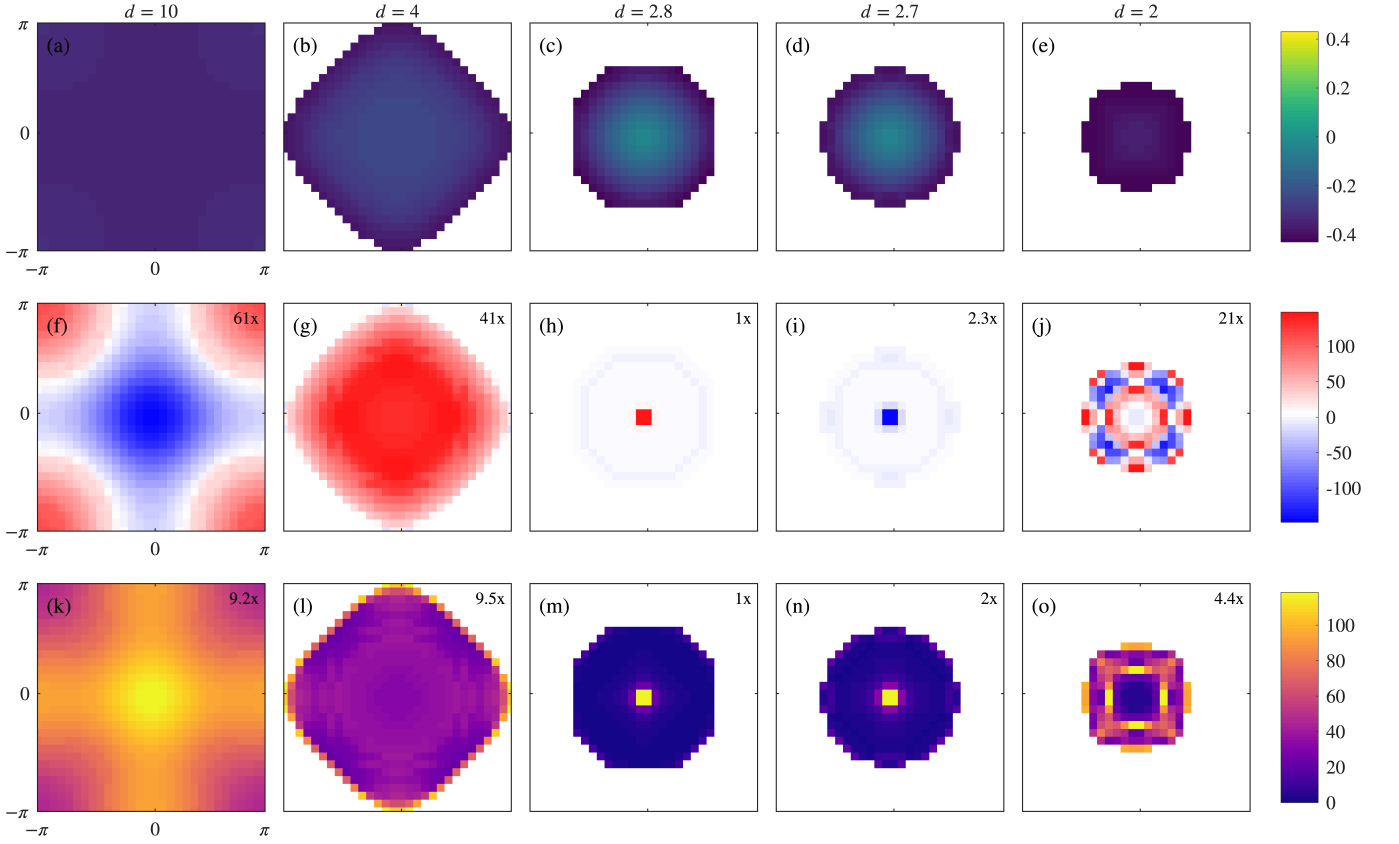}
\caption{Panels (a)--(e): evolution of the lower in-gap band structure across the mini Brillouin zone.
Panels (f)--(j): evolution of the Berry curvature $\Omega(\boldsymbol{\kappa})$ of the lower in-gap band.
Panels (k)--(o): evolution of $Q(\boldsymbol{\kappa})$ of the lower in-gap band.
Columns correspond to $d = 10, 4, 2.8, 2.7, 2$ [$W = 0.01, 0.063, 0.128, 0.137, 0.25$].
The numbers printed in the Berry-curvature and $Q$ panels indicate the per-panel color-scale zoom factor relative to the shared colorbar; 1x means the shared colorbar is used directly, while larger factors denote a zoomed-in color scale to make weaker features visible.
We use $m = 3.5$, $t = 2$, $v = 0.8$, $\Delta\simeq0.4622$, $\epsilon_b\simeq0.3373$, and $R_{\rm cut}=10.0$.
}
\label{fig:band_geometry_wf_ingapH}
\end{figure}

We can also investigate the band structure of the in-gap states, even though at different energies the spectrum is governed by a different effective Hamiltonian.
More specifically, for each $\boldsymbol{\kappa}$ we find the energy where $\epsilon \in h'(\boldsymbol{\kappa}, d, \epsilon)$ has a solution.
This allows us to plot the band structure in the mBZ, as shown in Fig.~\ref{fig:band_geometry_wf_ingapH}.
As $W$ becomes larger, for some momenta $\boldsymbol{\kappa}$ around the mBZ corner, an in-gap solution does not exist, producing the dark region around the $M$ point.

The in-gap Bloch wavefunctions are obtained from the many-impurity ring-state construction in SI~\ref{sup:many_impurity_green}, Eq.~\eqref{eq:ingap_wavefunction}, which expresses $\psi(\mathbf{r};\epsilon_m)$ as a superposition of single-vacancy ring-state Green's functions.
We write the Bloch form
\begin{equation}
\psi_{\boldsymbol{\kappa}}(\mathbf{r})=e^{i \boldsymbol{\kappa} \cdot \mathbf{r}} u_{\boldsymbol{\kappa}}(\mathbf{r}),
\end{equation}
with $u_{\boldsymbol{\kappa}}(\mathbf{r})$ cell periodic.
The Berry curvature and quantum metric are computed directly from this wavefunction via
\begin{equation}
\Omega(\boldsymbol{\kappa})=i\left(\langle \partial_{\kappa_x}u_{\boldsymbol{\kappa}}|\partial_{\kappa_y}u_{\boldsymbol{\kappa}}\rangle-\langle \partial_{\kappa_y}u_{\boldsymbol{\kappa}}|\partial_{\kappa_x}u_{\boldsymbol{\kappa}}\rangle\right),
\end{equation}
\begin{equation}
g_{ij}(\boldsymbol{\kappa})=\operatorname{Re}\left\langle\partial_{\kappa_i} u_{\boldsymbol{\kappa}}\right|(1-|u_{\boldsymbol{\kappa}}\rangle\langle u_{\boldsymbol{\kappa}}|)\left|\partial_{\kappa_j} u_{\boldsymbol{\kappa}}\right\rangle.
\end{equation}
Fig.~\ref{fig:band_geometry_wf_ingapH} summarizes these results.
The first row shows the lower in-gap band across the mBZ; the span of $\boldsymbol{\kappa}$ points where this band exists directly indicates the range in the BZ where an in-gap solution is present, and this range shrinks near the mBZ corners as $W$ increases.
Over the region where the physical in-gap solution exists, the second row plots $\Omega(\boldsymbol{\kappa})$ computed from $u_{\boldsymbol{\kappa}}$ and shows Berry curvature concentrating at $\Gamma$. The solution merges with the continuum near the mini-BZ boundary.
 The third row shows the spread measure $Q(\boldsymbol{\kappa})$, which diverges at $\Gamma$ because the band touching occurs at the $\Gamma$ point.

\begin{figure}[tbp]
\centering
\includegraphics[width=1.0\columnwidth]{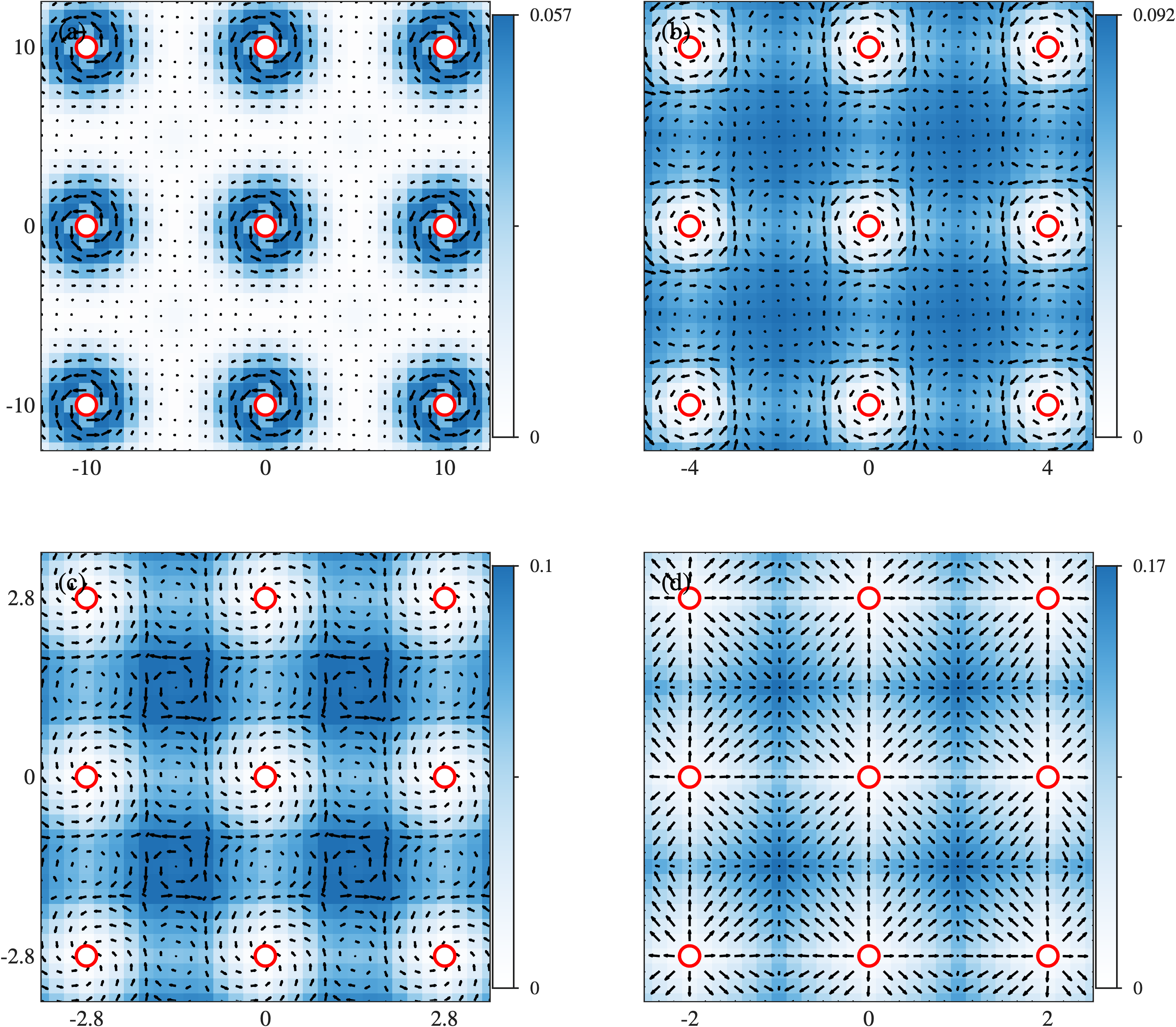}
\caption{$\Gamma$-point wavefunction in a periodic square vacancy array, shown in a central crop. Color denotes $|\psi(\mathbf{r})|^2$, arrows denote the plotted phase-gradient texture, and red circles mark vacancy positions.
Panels (a)--(d) correspond to $d = 10, 4, 2.7, 2$ [$W = 0.01, 0.063, 0.137, 0.25$], with $m = 3.5$, $t = 2$, and $v = 0.8$.
As $d$ decreases, the wavefunction density spreads across neighboring vacancies.}
\label{fig:in_gap_wf}
\end{figure}

We also visualize the $\Gamma$-point wavefunction for an open-boundary (vacancy) system. The conserved QWZ lattice current of a single state is defined on an oriented bond by
\begin{equation}
j_{\mathbf r\rightarrow\mathbf r'}
=-\frac{2}{\hbar}\operatorname{Im}
\!\left[\psi^\dagger_{\mathbf r}h_{\mathbf r\mathbf r'}\psi_{\mathbf r'}\right],
\end{equation}
where reversing the orientation reverses the sign. The arrows shown in Fig.~\ref{fig:in_gap_wf} instead plot $\operatorname{Im}[\psi^\dagger\nabla\psi]$.
Fig.~\ref{fig:in_gap_wf} shows the evolution of the $\Gamma$-point wavefunction computed for a large periodic square vacancy array. The central crop makes the probability-density and phase-gradient textures visible and shows increased wavefunction overlap as $d$ decreases.

\section{Berry curvature and quantum geometry evolution of the effective Hamiltonian}\label{sup:berry_geometry}

\begin{figure}[tbp]
\centering
\includegraphics[width=1.0\columnwidth]{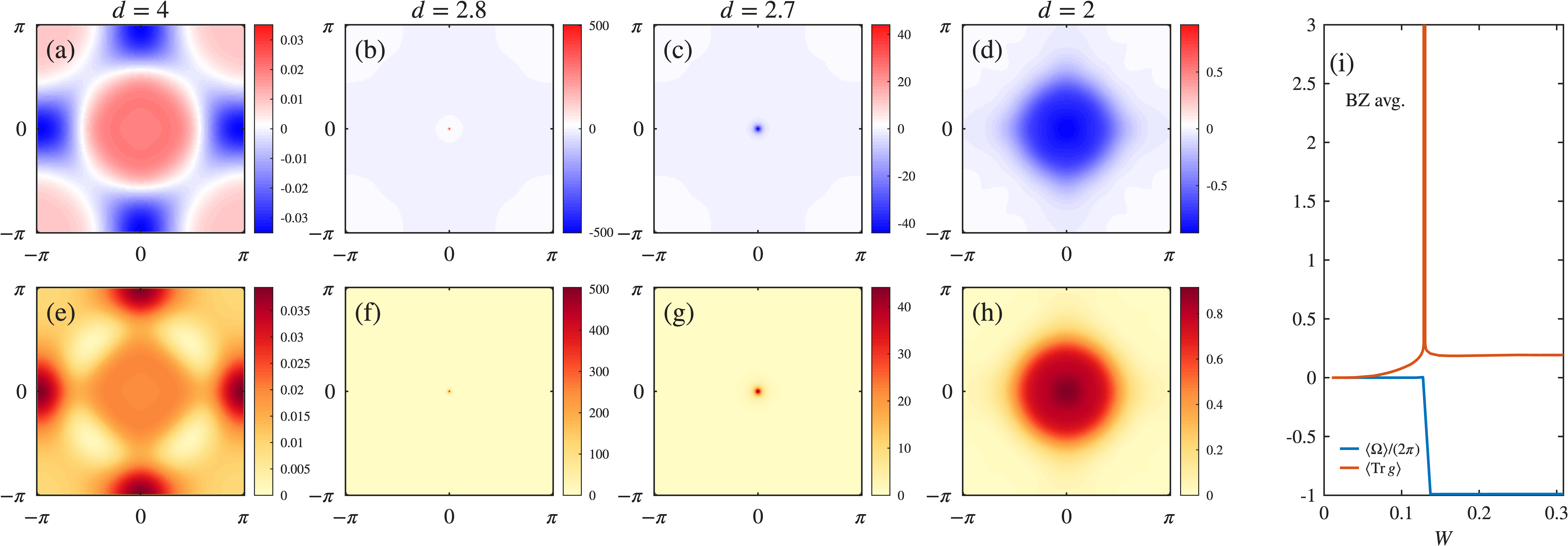}
\caption{Berry curvature and quantum geometry of the zero-energy effective Hamiltonian $h'(\epsilon = 0, \boldsymbol{\kappa})$.
Panels (a)--(d) show $\Omega(\boldsymbol{\kappa})$, and panels (e)--(h) show $\operatorname{Tr}g(\boldsymbol{\kappa})$, for $d = 4, 2.8, 2.7, 2$ [$W = 0.063, 0.128, 0.137, 0.25$] in the mini BZ (mBZ).
Panel (i) shows the mini-BZ integral of the lower-band Berry curvature of $h'(0)$ and the averaged metric trace; the former changes by one unit while the latter peaks near the closing.
We use $m = 3.5, t = 2, v = 0.8$.
}
\label{fig:Berry_geometry_H0}
\end{figure}

In order to see the key physics around zero energy, we do not need the full energy-dependent formalism; it is a good approximation to analyze the effective Hamiltonian at zero energy, $h'(\epsilon = 0, \boldsymbol{\kappa})$.
We visualize the evolution of Berry curvature and quantum geometry of $h'(\epsilon = 0, \boldsymbol{\kappa})$ in the mini BZ, as shown in Fig.~\ref{fig:Berry_geometry_H0}.
For the two-band model we are considering, $h'(\epsilon = 0, \boldsymbol{\kappa}) =\mathbf{d}_0(\boldsymbol{\kappa}) \mathbb{I}+\mathbf{d}(\boldsymbol{\kappa}) \cdot \boldsymbol{\sigma}$.
For the undeformed strict-vacancy kernel at $\epsilon=0$, $\mathbf d_0(\boldsymbol\kappa)=0$.
For the lower band, the Berry curvature is given by
\begin{equation}
\Omega(\boldsymbol{\kappa})=\frac{1}{2} \hat{\mathbf{d}} \cdot\left(\partial_{\kappa_x} \hat{\mathbf{d}} \times \partial_{\kappa_y} \hat{\mathbf{d}}\right)
\end{equation}
The quantum metric is the real part of the quantum geometric tensor
\begin{equation}
g_{i j}(\boldsymbol{\kappa})=\operatorname{Re}\left\langle\partial_{\kappa_i} u\right|(1-|u\rangle\langle u|)\left|\partial_{\kappa_j} u\right\rangle=\frac{1}{4} \partial_{\kappa_i} \hat{\mathbf{d}} \cdot \partial_{\kappa_j} \hat{\mathbf{d}}
\end{equation}
where the last equation holds for the two-band model.

At the Dirac band crossing point, the Berry curvature threads a relative flux of $-2\pi$ at the $\Gamma$ point in the effective description, accounting for the change of the total Chern number from $1$ to $0$.
Also because of the band touching, the quantum metric blows up at the transition point.
For the periodic-lattice data shown in the main text, the real-space hopping cutoff is $R_{\rm cut}=15$ for the spectral-window plot and $R_{\rm cut}=10.1$ for the Berry-curvature plot; the SI band-geometry figures use the cutoffs stated in their captions.

\section{Particle-hole asymmetric deformation}\label{sup:ph_asymmetry}

The undeformed clean Hamiltonian and the strict site-removal Hamiltonian obey an exact spectral PHS implemented by $\mathcal{C}=\sigma_xK$, with $K$ complex conjugation and $\mathcal{C}^2=+1$:
\begin{equation}
\mathcal{C}h_0(\bm k)\mathcal{C}^{-1}=-h_0(-\bm k).
\end{equation}
Literal removal of both orbitals at a vacancy preserves this relation and, with time reversal broken, places the strict-vacancy Hamiltonian globally in class D~\cite{AltlandZirnbauer1997}. At $E=0$ the symmetry constrains the same fixed-energy sector; at nonzero energy it relates the distinct $+E$ and $-E$ sectors, so a single fixed-energy localization problem has class-A constraints. A finite scalar defect $V_0\mathbb I$ breaks the spectral symmetry at every finite $V_0$.

To test the Green-function construction after explicitly breaking the spectral symmetry, we add the scalar deformation
\begin{equation}
h_\lambda(\bm k)=h_0(\bm k)+\lambda \bigl(2-\cos k_x-\cos k_y\bigr)\,\mathbb I .
\end{equation}
This term preserves translation symmetry and the square-lattice point group, while placing the full Hamiltonian in class A.
Near $\bm k=0$ it reduces to $\delta h_{\rm ph}(\bm k)\approx \frac{\lambda}{2}(k_x^2+k_y^2)\mathbb I$, so the Dirac mass at the gap minimum is unchanged but the dispersion away from the gap center becomes asymmetric.

\begin{figure}[tbp]
\centering
\includegraphics[width=1.0\columnwidth]{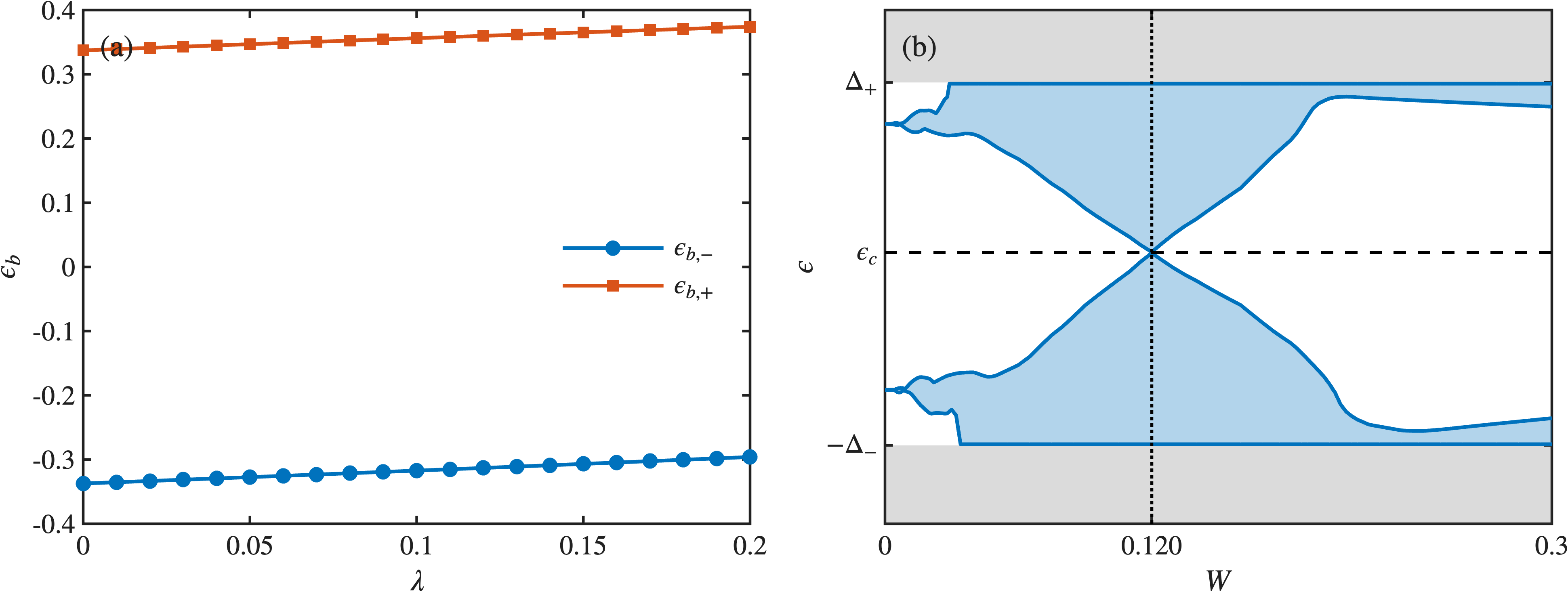}
\caption{Particle-hole asymmetric deformation of the vacancy problem.
(a) Single-vacancy bound-state energies $\epsilon_{b,\pm}$ versus deformation strength $\lambda$.
(b) In-gap solution window $S_\lambda$ at fixed $\lambda=0.2$ as a function of vacancy concentration $W$.
The blue region denotes impurity-band solutions and the gray region denotes the bulk continuum; dashed lines mark the finite-energy closing at $\epsilon_c\simeq 0.050$ and $W_c\simeq 0.120$.
For this slice, the clean bulk gap edges are $-\Delta_-\simeq -0.436$ and $\Delta_+\simeq 0.479$, and the single-vacancy energies are $\epsilon_{b,-}\simeq -0.296$ and $\epsilon_{b,+}\simeq 0.374$.
The transition remains captured by the same Green's-function construction after replacing $g_0\to g_{0,\lambda}$, but particle-hole asymmetry shifts the gap closing away from $\epsilon=0$.
Numerical parameters are $m=3.5$, $t=2$, $v=0.8$, $V_0=10^5$, and $R_{\rm cut}=10.1$.}
\label{fig:ph_asymmetry_summary}
\end{figure}

Numerically, we repeat exactly the same Green's-function construction as in the main text, now using the clean propagator of $h_\lambda$.
The Green-function pole construction does not rely on PHS: the bound-state energies still follow from $\det[g_{0,\lambda}(\epsilon_b;\bm a_0,\bm a_0)]=0$, and the in-gap solution window follows from the same self-consistency condition after replacing $g_0$ by $g_{0,\lambda}$.
The result is shown in Fig.~\ref{fig:ph_asymmetry_summary}.
The two bound-state branches shift unequally with $\lambda$, and the impurity-band window at fixed $\lambda=0.2$ remains continuously connected to the single-vacancy energies in the dilute limit while becoming visibly asymmetric about zero energy at larger vacancy concentration.
For comparison, the PHS periodic vacancy problem at the same $m=3.5$, $t=2$, and $v=0.8$ closes near $W_c\approx0.120$ at zero energy; the deformation shown here leaves the critical concentration near $W_c\simeq0.120$ but shifts the closing to the finite energy $\epsilon_c\simeq0.050$.

\section{Simplified transition condition}\label{sup:simplified_transition}

For the periodic vacancy lattice, the zero-energy transition can be obtained directly from the determinant condition in the main text, Eq.~\eqref{eq:det_condition}.
At zero energy we define the matrix $\mathcal{M}$ from the clean Green's function evaluated between vacancy sites.
Because the impurity lattice is translationally invariant, $\mathcal{M}$ is diagonalized by Bloch waves and the determinant factorizes into momentum sectors.
As discussed in the main text and seen numerically, the gap closes at the $\Gamma$ point, so it is enough to impose the transition condition in the $\boldsymbol{\kappa}=0$ sector.
Writing the zero-energy Green's function in momentum space and applying Poisson resummation for a square impurity lattice of spacing $d$, the $\Gamma$-point condition reduces to
\begin{equation}
\sum_{\mathbf{G}} h^{-1}(\mathbf{G}) = 0.
\end{equation}

Under the continuum approximation, Eq.~(S3) gives
\begin{equation}
h^{-1}(\mathbf{k})
=
\frac{1}{\xi_{\mathbf{k}}^2+\Delta^2}
\begin{pmatrix}
\xi_{\mathbf{k}} & \Delta e^{+i\phi_{\mathbf{k}}} \\
\Delta e^{-i\phi_{\mathbf{k}}} & -\xi_{\mathbf{k}}
\end{pmatrix},
\qquad
\xi_{\mathbf{k}}=k^2-\mu_\Gamma.
\end{equation}
Evaluated on the reciprocal lattice, the off-diagonal terms carry opposite angular phases.
Because the reciprocal lattice is inversion symmetric, these phases cancel in the sum over $\mathbf{G}$, and only the diagonal $\sigma_z$ component survives:
\begin{equation}
\sum_{\mathbf{G}} h^{-1}(\mathbf{G})
=
\left[
\sum_{\mathbf{G}}
\frac{\xi_{\mathbf{G}}}{\xi_{\mathbf{G}}^2+\Delta^2}
\right]\sigma_z,
\qquad
\xi_{\mathbf{G}}=G^2-\mu_\Gamma.
\end{equation}
The transition condition therefore becomes the scalar equation
\begin{equation}
S(d)\equiv
\sum_{m,n}
\frac{\left(\frac{2\pi}{d}\right)^2(m^2+n^2)-\mu_\Gamma}
{\left[\left(\frac{2\pi}{d}\right)^2(m^2+n^2)-\mu_\Gamma\right]^2+\Delta^2}
=
0,
\end{equation}
with the same ultraviolet cutoff as in the continuum Green's-function calculation.

Defining $\alpha \equiv (2\pi/d)^2$, $z \equiv (\mu_\Gamma+i\Delta)/\alpha$, and using
$x/(x^2+\Delta^2)=\operatorname{Re}[1/(x-i\Delta)]$, we obtain the compact form
\begin{equation}
S(d)
=
\frac{1}{\alpha}\operatorname{Re}
\sum_{m,n}
\frac{1}{m^2+n^2-z}.
\end{equation}
The transition point is determined by the root condition $S(d_c)=0$.
Even in this compact form, the dependence on $d$ remains nontrivial because it enters through both $\alpha=(2\pi/d)^2$ and the complex shift $z=(\mu_\Gamma+i\Delta)/\alpha$ inside an infinite two-dimensional lattice sum.
As a result, the transition condition remains implicit: in general it does not reduce to an elementary closed-form expression for $d_c$, and one must solve $S(d)=0$ numerically.

\section{Lower bound from quantum geometry and the flat-band limit}\label{sup:metric_bound_multiband}
\label{sec:metric_bound_multiband}

The main text uses the quantum geometry of the zero-energy effective Hamiltonian $h'(\epsilon=0,\boldsymbol{\kappa})$ as a diagnostic of the vacancy-band inversion.
At the transition this is an ordinary two-band Bloch Hamiltonian whose quantum-metric trace diverges when the Dirac mass closes, as shown in Fig.~\ref{fig:Berry_geometry_H0}(i).
This statement should be distinguished from a universal lower bound on the spatial extent of the full Green's-function bound state.
The energy-dependent Green's-function formalism does not define a single isolated-band projector for the dispersive impurity problem, and the flat-band metric bound derived below does not survive unchanged once band dispersion is included.

Consider the state generated by the full Green's function acting on a localized state,
\begin{equation}
|\psi(\mathbf k)\rangle
=
G(E,\mathbf k)|s\rangle
=
\sum_\alpha a_\alpha(\mathbf k)\,P_\alpha(\mathbf k)|s\rangle,
\qquad
a_\alpha(\mathbf k)=\frac{1}{E-\varepsilon_\alpha(\mathbf k)},
\end{equation}
where $P_\alpha(\mathbf k)=|u_\alpha(\mathbf k)\rangle\langle u_\alpha(\mathbf k)|$ is the projector onto band $\alpha$, $s_\alpha(\mathbf k)=\langle u_\alpha(\mathbf k)|s\rangle$, and therefore $|\psi(\mathbf k)\rangle=\sum_\alpha a_\alpha(\mathbf k)s_\alpha(\mathbf k)|u_\alpha(\mathbf k)\rangle$.
The second moment along $x$ is $\langle x^2\rangle=\int d^dk\, \|\partial_{k_x}\psi(\mathbf k)\|^2 / \int d^dk\, \|\psi(\mathbf k)\|^2$.
Writing $A^{(x)}_{\beta\alpha}(\mathbf k)=\langle u_\beta(\mathbf k)|\partial_{k_x}u_\alpha(\mathbf k)\rangle$ and using orthonormality of the Bloch basis, one finds
\begin{equation}
\partial_{k_x}|\psi\rangle
=
\sum_\beta
\left[
\partial_{k_x}(a_\beta s_\beta)
+
\sum_\alpha a_\alpha s_\alpha A^{(x)}_{\beta\alpha}
\right]
|u_\beta\rangle .
\end{equation}
Therefore
\begin{equation}
\|\partial_{k_x}\psi\|^2
=
\sum_\beta
\left|
\partial_{k_x}(a_\beta s_\beta)
+
\sum_\alpha a_\alpha s_\alpha A^{(x)}_{\beta\alpha}
\right|^2 =
\sum_\beta
\left|
\sum_{\alpha\neq\beta}
(a_\alpha-a_\beta)\,s_\alpha\,A^{(i)}_{\beta\alpha}
\right|^2 .
\label{eq:full_multiband_norm}
\end{equation}
We have used that $\partial_{k_i}s_\beta + \sum_\alpha A^{(i)}_{\beta\alpha}s_\alpha = 0$.
{The first equality is exact; the reduction to the second uses $a_\beta$ $\mathbf k$-independent, i.e.\ it holds for flat bands (the dispersive case is treated below).}
In general, the different terms interfere, so there is no universal lower bound depending only on the quantum metric for the full coherent multiband state.

\subsection{Multiband flat case}

We now assume each band is flat, though different bands may lie at different energies.
Then
$a_\alpha = 1/(E-\varepsilon_\alpha)$ is $\mathbf k$-independent, and the state can be written as
$|\psi(\mathbf k)\rangle = \sum_\alpha a_\alpha P_\alpha(\mathbf k)|s\rangle
= \sum_\alpha a_\alpha s_\alpha(\mathbf k)|u_\alpha(\mathbf k)\rangle$.

Now let $P(\mathbf k)=\sum_{n\in \mathrm{occ}} P_n(\mathbf k)$ project onto the bands below the gap containing $E$, and let $Q=1-P$.
If the occupied bands are degenerate at energy $\varepsilon_P$ and the empty bands at $\varepsilon_Q$, then
$G(E,\mathbf k)=a_P P(\mathbf k)+a_Q Q(\mathbf k)$ with
$a_P = 1/(E-\varepsilon_P)$ and $a_Q = 1/(E-\varepsilon_Q)$.
Hence $|\psi(\mathbf k)\rangle = a_P P(\mathbf k)|s\rangle + a_Q Q(\mathbf k)|s\rangle$ and
$\partial_{k_i}|\psi\rangle = (a_P-a_Q)(\partial_{k_i}P)|s\rangle$.
Using $P(\partial_{k_i}P)P=Q(\partial_{k_i}P)Q=0$, so that
$(\partial_{k_i}P)^2 = P(\partial_{k_i}P)Q(\partial_{k_i}P)P + Q(\partial_{k_i}P)P(\partial_{k_i}P)Q$,
we obtain
\begin{align}
\|\partial_{k_i}\psi\|^2
&=
|a_P-a_Q|^2
\Big[
\langle s|P(\partial_{k_i}P)Q(\partial_{k_i}P)P|s\rangle
\nonumber\\
&\qquad\qquad\qquad\qquad
+
\langle s|Q(\partial_{k_i}P)P(\partial_{k_i}P)Q|s\rangle
\Big] .
\label{eq:multiband_flat_projector}
\end{align}
Both terms are nonnegative.
This yields the lower bound
\begin{equation}
\|\partial_{k_i}\psi\|^2
\ge
|a_P-a_Q|^2\,
\langle s|P(\partial_{k_i}P)Q(\partial_{k_i}P)P|s\rangle = |a_P-a_Q|^2\ \mathfrak{g}^{ss}_{ii}(\mathbf k),
\label{eq:multiband_flat_bound}
\end{equation}
and similarly with $P\leftrightarrow Q$, where the operator $\mathfrak{g}_{ij}(\mathbf k)=P(\partial_{k_i}P)Q(\partial_{k_j}P)P$
is the operator-valued quantum metric on the occupied subspace, with matrix elements $\mathfrak{g}^{ss}_{ij}(\mathbf k)=\langle s|P(\partial_{k_i}P)Q(\partial_{k_j}P)P|s\rangle$ for a given source $|s\rangle$.
Equation~\eqref{eq:multiband_flat_bound} therefore shows that, in the multiband flat case, the spatial spread admits an exact lower bound controlled by the quantum geometry of the occupied projector.

This bound still depends on the choice of $|s\rangle$.
The total Chern number controls only the trace of the integrated metric,
$\int d^2k\, \Tr g^P(\mathbf k) \ge 2\pi |C|$,
and does not in general bound each matrix element
$
\mathfrak{g}^{ss}_{ij}(\mathbf k)
$
from below.
Thus we obtain a geometric lower bound for a fixed $|s\rangle$, but not a universal one in terms of the Chern number alone.
However, if $\{|e_a\rangle\}$ is a complete orthonormal basis of source orbitals, then summing Eq.~\eqref{eq:multiband_flat_bound} over $a$ gives
\begin{equation}
\sum_a \sum_{i=x,y}\|\partial_{k_i}\psi_{e_a}\|^2
\ge
|a_P-a_Q|^2 \sum_a \sum_{i=x,y} \mathfrak{g}^{e_a e_a}_{ii}(\mathbf k)
=
|a_P-a_Q|^2 \Tr g^P(\mathbf k),
\end{equation}
and therefore, after integrating over the Brillouin zone,
\begin{equation}
\sum_a \int d^2k\, \sum_{i=x,y}\|\partial_{k_i}\psi_{e_a}\|^2
\ge
|a_P-a_Q|^2 \int d^2k\, \Tr g^P(\mathbf k)
\ge
2\pi |C|\, |a_P-a_Q|^2 .
\end{equation}
Since each term is nonnegative, in a Chern band not all source-orbital spreads can vanish simultaneously.

\subsection{Two-band case}

Now restrict to two bands.
Eq.~\eqref{eq:full_multiband_norm} becomes
In the flat-band limit the coefficients $a_\alpha$ become independent of $\mathbf k$.
Since the source $|s\rangle$ is $k$-independent, the overlaps
\begin{equation}
\|\partial_{k_x}\psi\|^2
=
|a_1-a_2|^2
\left(
|s_1|^2 |A^{(x)}_{21}|^2
+
|s_2|^2 |A^{(x)}_{12}|^2
\right).
\end{equation}
For a two-band model,
$
|A^{(x)}_{21}|^2 = |A^{(x)}_{12}|^2 = g_{xx}(\mathbf k),
$
so, using $|s_1|^2+|s_2|^2=\langle s|s\rangle$ for a normalized source, we obtain
\begin{equation}
\|\partial_{k_x}\psi(\mathbf k)\|^2
=
|a_1-a_2|^2\, g_{xx}(\mathbf k)\,\big(|s_1(\mathbf k)|^2+|s_2(\mathbf k)|^2\big).
\end{equation}
where $g_{xx}$ is the quantum metric of the two-band problem, namely $g_{xx}(\mathbf k)=\langle \partial_{k_x}u_1|(1-P_1)|\partial_{k_x}u_1\rangle=|A^{(x)}_{21}|^2$.
Let $\mathcal N(E)$ denote the normalization factor of the state.
Then
\begin{equation}
\langle x^2\rangle = \frac{|\varepsilon_1-\varepsilon_2|^2}{|E-\varepsilon_1|^2 |E-\varepsilon_2|^2}\,\mathcal N(E)\int d^dk\, g_{xx}(\mathbf k).
\end{equation}
For flat bands we also have
\begin{equation}
\mathcal N(E)^{-1} = \int d^2k\,\left(\frac{|s_1(\mathbf k)|^2}{|E-\varepsilon_1|^2} + \frac{|s_2(\mathbf k)|^2}{|E-\varepsilon_2|^2}\right) = \frac{S_1}{|E-\varepsilon_1|^2} + \frac{S_2}{|E-\varepsilon_2|^2},
\end{equation}
where
$
S_\alpha=\int d^2k\,|s_\alpha(\mathbf k)|^2
$
and $S_1+S_2=V_{\rm BZ}$ for a normalized source.
Therefore
\begin{equation}
\frac{|\varepsilon_1-\varepsilon_2|^2}{|E-\varepsilon_1|^2 |E-\varepsilon_2|^2}\,\mathcal N(E)
=
\frac{|\varepsilon_1-\varepsilon_2|^2}{S_1|E-\varepsilon_2|^2+S_2|E-\varepsilon_1|^2}
\ge
\frac{1}{\max(S_1,S_2)}
\ge
\frac{1}{V_{\rm BZ}},
\end{equation}
since for $E$ inside the gap the denominator is bounded by $\max(S_1,S_2)|\varepsilon_1-\varepsilon_2|^2$.
Thus, for two flat bands, the spatial spread is controlled entirely by the quantum geometry.
In two dimensions, summing the second moments along $x$ and $y$ gives
\begin{equation}
\langle x^2+y^2\rangle = \frac{|\varepsilon_1-\varepsilon_2|^2}{|E-\varepsilon_1|^2 |E-\varepsilon_2|^2}\,\mathcal N(E)\int d^2k\, \Tr g(\mathbf k) \ge \frac{2\pi |C|\,|\varepsilon_1-\varepsilon_2|^2}{|E-\varepsilon_1|^2 |E-\varepsilon_2|^2}\,\mathcal N(E) \ge \frac{2\pi |C|}{V_{\rm BZ}} .
\end{equation}
where $C$ is the Chern number, and the bound follows from $\Tr g \ge |\Omega|$ together with $\int d^2k\,|\Omega| \ge 2\pi |C|$.
For the standard Brillouin-zone volume $V_{\rm BZ}=(2\pi)^2$, this becomes
\begin{equation}
\langle x^2+y^2\rangle \ge \frac{|C|}{2\pi}.
\end{equation}

\begin{figure}[tbp]
\makebox[\columnwidth][c]{\includegraphics[width=0.8\columnwidth]{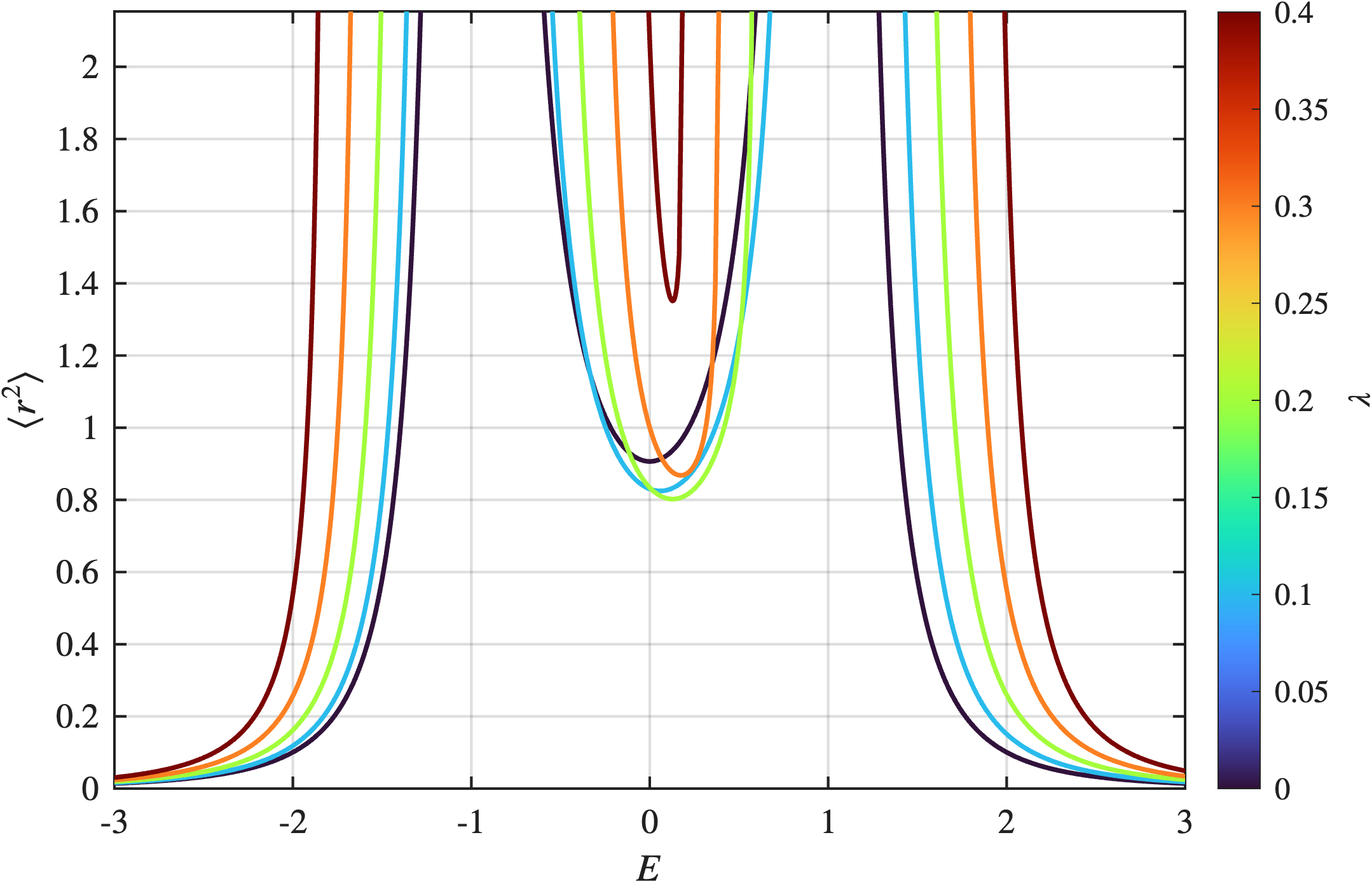}}
\caption{Numerical experiment for the non-flat two-band case using fixed QWZ eigenvectors and dispersive band energies $\varepsilon_\pm(\mathbf k)=\pm \Delta_0 \pm \lambda(\cos k_x+\cos k_y)$ with $\Delta_0=1$.
The horizontal axis is the probe energy $E$, while the vertical axis is labeled by the spread measure $\langle r^2\rangle$.
 Since the eigenvectors are unchanged, the quantum metric is held fixed throughout the scan; changing $\lambda$ changes both the resolvent weights and their momentum derivatives. The quantity defined in the text is the unnormalized derivative norm $F(E,\lambda)$, rather than the normalized spread.}
\label{fig:second_moment_dispersion}
\end{figure}

For comparison, in the non-flat two-band case $a_\alpha(\mathbf k)=1/(E-\varepsilon_\alpha(\mathbf k))$ is momentum dependent, so the analogue of Eq.~\eqref{eq:multiband_flat_bound} is no longer purely geometric.
Instead we find
\begin{equation}
\|\partial_{k_x}\psi\|^2
=
\left|
(\partial_{k_x}a_1)\,s_1+(a_2-a_1)\,s_2\,A^{(x)}_{12}
\right|^2
+
\left|
(\partial_{k_x}a_2)\,s_2+(a_1-a_2)\,s_1\,A^{(x)}_{21}
\right|^2 .
\end{equation}
Using $\partial_{k_x}a_\alpha=(\partial_{k_x}\varepsilon_\alpha)/(E-\varepsilon_\alpha)^2$, this can be expanded as
\begin{align}
\|\partial_{k_x}\psi\|^2
&=
|a_1-a_2|^2\, g_{xx}(\mathbf k)
+ |\partial_{k_x}a_1|^2 |s_1|^2
+ |\partial_{k_x}a_2|^2 |s_2|^2
\nonumber\\
&\quad
+ 2\,\Re\!\left[
(\partial_{k_x}a_1)^*(a_2-a_1)s_1^*s_2A^{(x)}_{12}
+ (\partial_{k_x}a_2)^*(a_1-a_2)s_2^*s_1A^{(x)}_{21}
\right].
\end{align}
Thus band dispersion contributes additional positive terms as well as interference terms with the quantum-geometric piece, so the simple flat-band relation between spatial spread and metric no longer holds in general.
To illustrate this breakdown numerically, we consider a two-band model in which the Bloch eigenvectors are kept fixed while a tunable dispersion is added only to the band energies.
The fixed eigenvectors are those of the QWZ Hamiltonian
$
H_{\rm geom}(\mathbf k)=\mathbf d(\mathbf k)\cdot \boldsymbol{\sigma},
\qquad
\mathbf d(\mathbf k)=\big(\sin k_x,\sin k_y,m+\cos k_x+\cos k_y\big),
$
with $m=-1$.
On top of this fixed geometry we impose the dispersive band energies
$
\varepsilon_\pm(\mathbf k)=\pm \Delta_0 \pm \lambda(\cos k_x+\cos k_y).
$
In this construction the quantum geometry is unchanged as $\lambda$ varies, whereas the new contribution enters entirely through $\partial_{k_i}a_\alpha(\mathbf k)$.
Figure~\ref{fig:second_moment_dispersion} shows the Brillouin-zone-averaged derivative norm
$
F(E,\lambda)=\langle \|\partial_{k_x}\psi_E\|^2+\|\partial_{k_y}\psi_E\|^2\rangle_{\mathrm{BZ}},
$
which in the flat-band limit directly controls the spatial spread.
The flat case $\lambda=0$ follows the purely geometric behavior, while increasing $\lambda$ produces visible deviations, especially away from the middle of the gap, consistent with the extra dispersion and interference terms in the expression above.

 The dispersive calculation no longer obeys the flat-band equality between the derivative norm and the quantum metric. Figure~\ref{fig:second_moment_dispersion} does not, as presently defined, establish a violation of the normalized bound $\langle x^2+y^2\rangle\ge2\pi|C|/V_{\rm BZ}$. Such a test requires normalization of the generated state and a probe energy inside the central gap, $|E|<1-2|\lambda|$, for the parameters used here.

\section{Self-consistent Born approximation for positional disorder}
\label{sup:kernel_SCBA}

Here we give the disorder-averaged effective-medium construction used in the main text and its range of validity.
For fixed physical energy $\epsilon$ entering the vacancy Hamiltonian $h'(\epsilon)$, we introduce an auxiliary spectral variable $\tilde\epsilon$ and define
\begin{align}
\overline{g}^{-1}(\tilde\epsilon,\epsilon,\boldsymbol{\kappa})
&=
\tilde\epsilon+i0^+-h'_0(\epsilon,\boldsymbol{\kappa})
-\Sigma(\tilde\epsilon,\epsilon,\boldsymbol{\kappa}),\\
h'_{\rm top}(\epsilon,\boldsymbol{\kappa})
&=
h'_0(\epsilon,\boldsymbol{\kappa})
+\mathrm{Re}\,\Sigma(\tilde\epsilon=\epsilon,\epsilon,\boldsymbol{\kappa}).
\end{align}
This is the topological-Hamiltonian construction applied to the disorder-averaged ring-state Green's function~\cite{volovik2003universe,gurarie2011single,wang2010topological,wang2012simplified,groth2009theory}.
We use $h'_{\rm top}$ as an effective-medium way to locate the disorder-renormalized mass inversion, not as the exact sample-resolved Chern invariant of the full disorder ensemble.
In general, computing a Chern number after disorder averaging is not equivalent to computing the invariant in each sample and then averaging.
 Vanishing $\mathrm{Im}\,\Sigma(\tilde\epsilon=\epsilon)$ is an additional effective-medium assumption, appropriate to a spectral gap without elastic spectral weight; a mobility gap alone does not imply it. The construction also assumes that the averaged Green's function has no zero-energy zeros. These assumptions do not establish equality with the sample-resolved invariant at the localization transition.

The disorder-averaged self-energy in the SCBA satisfies \cite{bruus2004many}
\begin{equation}
\Sigma_{\alpha\beta}(\tilde{\epsilon}, \epsilon, \boldsymbol{\kappa})
=  \sum_{\boldsymbol{\kappa}', \gamma, \delta}
\mathcal{C}_{\alpha\gamma;\delta\beta}(\epsilon; \boldsymbol{\kappa}, \boldsymbol{\kappa}')
\, \bar{g}^{\gamma\delta}(\tilde{\epsilon}, \epsilon, \boldsymbol{\kappa}'),
\end{equation}
where the momentum sum carries no additional $1/N$, because $\delta h'(\boldsymbol{\kappa},\boldsymbol{\kappa}')$ below is defined as a matrix element between normalized Bloch states and the kernel $\mathcal{C}$ is correspondingly $O(1/N)$, and
$
\bar{g}(\tilde{\epsilon}, \epsilon, \boldsymbol{\kappa})
= \big[\tilde{\epsilon} + i0^+ - {h}'_0(\epsilon, \boldsymbol{\kappa}) - \Sigma(\tilde{\epsilon}, \epsilon, \boldsymbol{\kappa})\big]^{-1}
$
is the disorder-averaged Green’s function.
The disorder kernel is defined by
\begin{equation}
\mathcal{C}_{\alpha\gamma;\delta\beta}(\epsilon, \boldsymbol{\kappa}, \boldsymbol{\kappa}')
=
\big\langle
\delta {h'}^{\alpha\gamma}(\epsilon, \boldsymbol{\kappa}, \boldsymbol{\kappa}')\,
\delta {h'}^{\delta\beta}(\epsilon, \boldsymbol{\kappa}', \boldsymbol{\kappa})
\big\rangle_{\mathrm{dis}}.
\end{equation}
Below we suppress the fixed $\epsilon$ argument in $\mathcal{C}$ and $\delta h'$ for compactness.
We consider weak positional fluctuations of the vacancy sites
\(\mathbf{a}_i = \bar{\mathbf{a}}_i + \boldsymbol{\delta}_i\),
where \(\bar{\mathbf{a}}_i\) are the ideal lattice sites and
\(\boldsymbol{\delta}_i\) are small random displacements.
Expanding the effective hopping matrix between vacancy orbitals to first order in \(\boldsymbol{\delta}_i\) gives
\begin{equation}
\delta {h}'^{\alpha\beta}(\mathbf{a}_i - \mathbf{a}_j)
\simeq
\nabla_\mu{h}'^{\alpha\beta}(\bar{\mathbf{a}}_{ij})
(\delta_i^\mu - \delta_j^\mu),
\qquad
\bar{\mathbf{a}}_{ij} \equiv \bar{\mathbf{a}}_i - \bar{\mathbf{a}}_j,
\end{equation}
where the repeated Cartesian index \(\mu\) is summed.
{At second order in the displacements the expansion contains $\tfrac{1}{2}\nabla_\mu\nabla_\nu h'^{\alpha\beta}(\bar{\mathbf{a}}_{ij})(\delta_i^\mu-\delta_j^\mu)(\delta_i^\nu-\delta_j^\nu)$, whose disorder average does not vanish: with the displacement statistics specified below, $\langle(\delta_i^\mu-\delta_j^\mu)(\delta_i^\nu-\delta_j^\nu)\rangle_{\mathrm{dis}}=\sigma_u^2\,\delta^{\mu\nu}$ for $i\neq j$, so
\begin{equation}
\big\langle \delta {h}'^{\alpha\beta}(\mathbf{a}_i-\mathbf{a}_j)\big\rangle_{\mathrm{dis}}
=\frac{\sigma_u^2}{2}\,\nabla^2 {h}'^{\alpha\beta}(\bar{\mathbf{a}}_{ij})+O(\sigma_u^4).
\end{equation}
This Debye--Waller-type mean is $O(\sigma_u^2)$, the same order as the Born mass shift, and must be retained. We absorb it into the periodic reference Hamiltonian,
\begin{equation}
h'_0(\epsilon,\boldsymbol{\kappa})\;\to\;
h'_0(\epsilon,\boldsymbol{\kappa})
+\frac{\sigma_u^2}{2}\sum_{\bar{\mathbf{a}}\neq0}
e^{-i\boldsymbol{\kappa}\cdot\bar{\mathbf{a}}}\,
\nabla^2 h'(\bar{\mathbf{a}},\epsilon),
\end{equation}
and solve the SCBA below around this shifted reference, with the kernel built from the zero-mean fluctuation $\delta h'-\langle\delta h'\rangle_{\mathrm{dis}}$; the residual second-order fluctuation contributes to the kernel only at $O(\sigma_u^4)$.}

The fluctuation of the Bloch Hamiltonian matrix element is
\begin{equation}
\delta {h}'^{\alpha\beta}(\boldsymbol{\kappa}, \boldsymbol{\kappa}')
=
{\frac{1}{N}}\sum_{i,j}
e^{-i(\boldsymbol{\kappa}\cdot\bar{\mathbf{a}}_i - \boldsymbol{\kappa}'\cdot\bar{\mathbf{a}}_j)}\,
\delta {h}'^{\alpha\beta}(\mathbf{a}_i - \mathbf{a}_j).
\end{equation}
{i.e.\ the matrix element between normalized Bloch states $|\boldsymbol{\kappa}\rangle=N^{-1/2}\sum_i e^{i\boldsymbol{\kappa}\cdot\bar{\mathbf{a}}_i}|i\rangle$; this is the same normalization for which the periodic part reduces to $h'_0(\boldsymbol{\kappa},d,\epsilon)=\sum_{\mathbf{a}}h'(\mathbf{a},\epsilon)e^{-i\boldsymbol{\kappa}\cdot\mathbf{a}}$ of the main text.}
The disorder kernel entering the SCBA is defined as
\begin{equation}
\mathcal{C}_{\alpha\gamma;\delta\beta}(\boldsymbol{\kappa}, \boldsymbol{\kappa}')
=
\big\langle
\delta {h}'^{\alpha\gamma}(\boldsymbol{\kappa}, \boldsymbol{\kappa}')
\,\delta {h}'^{\delta\beta}(\boldsymbol{\kappa}', \boldsymbol{\kappa})
\big\rangle_{\mathrm{dis}}.
\end{equation}
where $\langle...\rangle_{\operatorname{dis}}$ denotes averaging over random displacements.
Assuming uncorrelated isotropic displacements
\begin{equation}
\langle \delta_i^\mu \delta_j^\nu \rangle_{\mathrm{dis}}
= {\frac{\sigma_u^2}{2}}\,\delta_{ij}\,\delta^{\mu\nu},
\end{equation}
where {$\sigma_u^2\equiv\langle|\boldsymbol{\delta}_i|^2\rangle$ is the total displacement variance, so that in two dimensions each Cartesian component carries variance $\sigma_u^2/2$; this is the same convention as the bimodal mapping of Eq.~\eqref{eq:bimodal_sigma_relation}, which fixes the trajectory $\sigma_u=1/\sqrt{6W}$},
we obtain
\begin{equation}
\mathcal{C}_{\alpha\gamma;\delta\beta}(\boldsymbol{\kappa}, \boldsymbol{\kappa}')
=
{\frac{\sigma_u^2}{2N^2}}
\sum_{i,j,i',j'}
e^{-i(\boldsymbol{\kappa}\cdot\bar{\mathbf{a}}_i - \boldsymbol{\kappa}'\cdot\bar{\mathbf{a}}_j
+ \boldsymbol{\kappa}'\cdot\bar{\mathbf{a}}_{i'} - \boldsymbol{\kappa}\cdot\bar{\mathbf{a}}_{j'})}
\nabla_\mu {h}'^{\alpha\gamma}(\bar{\mathbf{a}}_{ij})
\,\nabla_\mu {h}'^{\delta\beta}(\bar{\mathbf{a}}_{i'j'})
(\delta_{ii'} - \delta_{ij'} - \delta_{ji'} + \delta_{jj'}).
\end{equation}
Transforming the sums to bond vectors
\(\bar{\mathbf{a}} = \bar{\mathbf{a}}_{ij}\),
\(\bar{\mathbf{a}}' = \bar{\mathbf{a}}_{i'j'}\),
and performing the remaining site sums gives
\begin{align}
\mathcal{C}_{\alpha\gamma;\delta\beta}(\boldsymbol{\kappa}, \boldsymbol{\kappa}')
=
{\frac{\sigma_u^2}{2N}}
\sum_{\bar{\mathbf{a}},\bar{\mathbf{a}}'}
e^{-i(\boldsymbol{\kappa}'\cdot\bar{\mathbf{a}} + \boldsymbol{\kappa}\cdot\bar{\mathbf{a}}')}
(1 - e^{-i\mathbf{q}\cdot\bar{\mathbf{a}}'})
(1 - e^{i\mathbf{q}\cdot\bar{\mathbf{a}}})
\nabla {h}'^{\alpha\gamma}(\bar{\mathbf{a}})
\cdot
\nabla {h}'^{\delta\beta}(\bar{\mathbf{a}}'),
\qquad \mathbf{q} \equiv \boldsymbol{\kappa}' - \boldsymbol{\kappa}.
\end{align}
where $N$ denotes the number of impurities.
Define the vector-valued vertex function
\begin{equation}
\mathbf{T}_{\alpha\beta}(\boldsymbol{\kappa})
\equiv
\sum_{\bar{\mathbf{a}}}
e^{-i\boldsymbol{\kappa}\cdot\bar{\mathbf{a}}}
\nabla {h}'^{\alpha\beta}(\bar{\mathbf{a}}),
\end{equation}
for which
\begin{align}
\sum_{\bar{\mathbf{a}}}
e^{-i\boldsymbol{\kappa}'\cdot\bar{\mathbf{a}}}
(1 - e^{i\mathbf{q}\cdot\bar{\mathbf{a}}})
\nabla {h}'^{\alpha\gamma}(\bar{\mathbf{a}})
&=
\mathbf{T}_{\alpha\gamma}(\boldsymbol{\kappa}') - \mathbf{T}_{\alpha\gamma}(\boldsymbol{\kappa}),\\
\sum_{\bar{\mathbf{a}}'}
e^{-i\boldsymbol{\kappa}\cdot\bar{\mathbf{a}}'}
(1 - e^{-i\mathbf{q}\cdot\bar{\mathbf{a}}'})
\nabla {h}'^{\delta\beta}(\bar{\mathbf{a}}')
&=
\mathbf{T}_{\delta\beta}(\boldsymbol{\kappa}) - \mathbf{T}_{\delta\beta}(\boldsymbol{\kappa}').
\end{align}
Substituting these relations yields the compact form
\begin{equation}
    \label{eq:SCBA_kernel}
\mathcal{C}_{\alpha\gamma;\delta\beta}(\boldsymbol{\kappa},\boldsymbol{\kappa}')
=
{\frac{\sigma_u^2}{2N}}\,
[\mathbf{T}_{\alpha\gamma}(\boldsymbol{\kappa}') - \mathbf{T}_{\alpha\gamma}(\boldsymbol{\kappa})]
\cdot
[\mathbf{T}_{\delta\beta}(\boldsymbol{\kappa}) - \mathbf{T}_{\delta\beta}(\boldsymbol{\kappa}')]
.
\end{equation}
This is our central result.
{Substituting Eq.~\eqref{eq:SCBA_kernel} into the SCBA equation gives the manifestly intensive self-energy
\begin{equation}
\Sigma_{\alpha\beta}(\tilde{\epsilon},\epsilon,\boldsymbol{\kappa})
=\frac{\sigma_u^2}{2N}\sum_{\boldsymbol{\kappa}',\gamma,\delta}
[\mathbf{T}_{\alpha\gamma}(\boldsymbol{\kappa}')-\mathbf{T}_{\alpha\gamma}(\boldsymbol{\kappa})]\cdot
\bar{g}^{\gamma\delta}(\tilde{\epsilon},\epsilon,\boldsymbol{\kappa}')\,
[\mathbf{T}_{\delta\beta}(\boldsymbol{\kappa})-\mathbf{T}_{\delta\beta}(\boldsymbol{\kappa}')],
\end{equation}
which has a finite thermodynamic limit, since $\frac{1}{N}\sum_{\boldsymbol{\kappa}'}\to \frac{d^2}{(2\pi)^2}\int_{\rm mBZ}d^2\kappa'$.}This expression shows that the positional-disorder correlation depends on the momentum transfer
\(\mathbf{q} = \boldsymbol{\kappa}' - \boldsymbol{\kappa}\)
only through the difference of the vertex functions \(\mathbf{T}(\boldsymbol{\kappa})\).

The difference structure of the kernel follows from the disorder originating in \emph{relative} displacements of the impurities, and it controls the entire weak-disorder regime. Expanding for small momentum transfer, $\mathbf{T}(\boldsymbol{\kappa}')-\mathbf{T}(\boldsymbol{\kappa})\simeq q_i\,\partial_{\kappa_i}\mathbf{T}(\boldsymbol{\kappa})$, so that
\begin{equation}
\mathcal{C}(\boldsymbol{\kappa},\boldsymbol{\kappa}')\propto\sigma_u^2\,|\mathbf{q}|^2,
\qquad |\mathbf{q}|\rightarrow0 .
\end{equation}
This infrared structure is fixed by translational invariance: a uniform shift of all impurity positions leaves the Hamiltonian unchanged, so the disorder vertex must vanish at zero momentum transfer and forward scattering is suppressed. Near the zero-energy transition the Born-renormalized ring-state Hamiltonian is Dirac-like,
\begin{equation}
h'_{\rm Born}(\boldsymbol{\kappa})
=\tilde v\,(\kappa_x\sigma_x-\kappa_y\sigma_y)
+\bigl[\tilde m_{\rm Born}-\tilde t\,|\boldsymbol{\kappa}|^2\bigr]\sigma_z+\cdots,
\end{equation}
with the transition set by the real mass inversion $\tilde m_{\rm Born}(W,\sigma_u)=0$; the SCBA imaginary part vanishes in the gap because no elastic channel is available at $\epsilon=0$. The same $|\mathbf{q}|^2$ suppression controls the damping: on shell the scattering rate obeys $\Gamma(\omega)/|\omega|\propto|\omega|^2$, parametrically smaller than for momentum-independent disorder, so positional disorder produces an appreciable real mass renormalization while keeping the Dirac point sharp. At fourth order in a symmetric displacement distribution, corrections include crossed diagrams, the covariance of quadratic hopping fluctuations, linear--cubic terms, and the fourth-order mean. 
{ This expansion is in the vacancy displacements, not in the vacancy potential: each vacancy is a unitary scatterer treated exactly through the periodic reference Hamiltonian $h'_0$, and the expansion parameter is $\sigma_u$.} The displayed treatment includes the mean and Born terms through second order in the weak positional-disorder expansion {, with the $O(\sigma_u^2)$ Debye--Waller mean $\langle\delta h'\rangle_{\mathrm{dis}}=\frac{\sigma_u^2}{2}\nabla^2 h'$ included in the periodic reference Hamiltonian as described above ($\Sigma_{\rm num}$ contains it automatically, since it never expands in $\delta h'$)}. This conclusion applies only to the weakly displaced lattice and does not recover the independent bimodal transition.
 This control is asymptotic in $\sigma_u$. Along the heuristic trajectory $\sigma_u=1/\sqrt{6W}$ the disorder reaches $\sigma_u=O(1)$, outside the weak-displacement regime. Within the displaced-lattice ensemble the numerical disorder-averaged self-energy is then more reliable than the second-order SCBA, but neither calculation constitutes a controlled prediction for the { bimodal} ensemble.

\section{Numerical studies of the localization transition}\label{sup:localization_lcm}

\subsection{Bimodal disorder}

We numerically study the vacancy problem with a bimodal on-site potential,
\begin{equation}
\hat{V}_{\operatorname{bimodal}}
= \sum_{i\alpha} V_i c_{\mathbf{a}_i,\alpha}^{\dagger}c_{\mathbf{a}_i,\alpha},
\qquad
V_i =
\begin{cases}
V_0, & \text{probability} W,\\
0, & \text{probability} 1-W,
\end{cases}
\end{equation}
where $\alpha=s,p$ labels the two orbitals.
We take $V_0$ large, so a site with $V_i=V_0$ acts as a vacancy, and $W$ is the vacancy concentration. { For $V_0\to\infty$ the vacancies are unitary scatterers, so the disorder is strong at every impurity site and admits no expansion in the impurity potential.}

To compare this { bimodal} disorder heuristically with the positional-disorder SCBA calculation in the main text, we construct an impurity lattice with the same areal density and match the second moment of the impurity position within a cell.
On a square host lattice with unit lattice constant, the impurity density is $n_{\rm imp}=W$.
Replacing the random vacancies by one impurity per square impurity cell gives $d^{-2}=W$, where $d$ is the impurity-lattice spacing.
If the impurity position is uniformly distributed within this cell, then
\begin{equation}
\sigma_u^2 \equiv \langle |\boldsymbol{\delta}|^2\rangle
= \langle \delta_x^2+\delta_y^2\rangle
=\frac{d^2}{6}
=\frac{1}{6W}.
\label{eq:bimodal_sigma_relation}
\end{equation}
This gives the dotted trajectory $\sigma_u=1/\sqrt{6W}$ in Fig.~\ref{fig:random_percolation}(a){, in the total-variance convention for $\sigma_u$ used throughout}.
The construction matches only the density and this second moment; it does not make the disorder ensembles equivalent. In particular, independent occupations have a finite long-wavelength structure factor proportional to $W(1-W)$, whereas the randomly displaced lattice is hyperuniform and its structure factor vanishes as $q\to0$. Moreover, $\sigma_u\simeq1$ near the { bimodal} critical concentration, so this trajectory is outside the controlled weak-displacement regime of the SCBA. We therefore show it only as a heuristic guide and determine the { bimodal} transition directly from the full-lattice TM, local-Chern-marker, and multifractal calculations.
Any numerical proximity between the SCBA gap closing along this guide and the TM crossing should therefore not be interpreted as a controlled mapping between the two ensembles.

\subsection{Local Chern marker and current texture}\label{sup:local_chern_current}

The local Chern marker and arrow-texture maps in Fig.~\ref{fig:random_percolation}(d)--(f) give a real-space diagnostic of the same transition.
For these maps we use the standard real-space diagnostics~\cite{bianco2011mapping,caio2019topological,wang2024anderson}
\begin{equation}
C(\mathbf{r})=\frac{4\pi}{A_c}\sum_\alpha
\operatorname{Im}
\langle \mathbf{r}\alpha|\hat{P}\hat{r}_x\hat{P}\hat{r}_y\hat{P}|\mathbf{r}\alpha\rangle,
\qquad
q_i(\mathbf{r})=\sum_\alpha
\operatorname{Im}
\langle \mathbf{r}\alpha|\hat{J}_i\hat{P}|\mathbf{r}\alpha\rangle,
\qquad
j_{\mathbf r\to\mathbf r'}=
-\frac{2}{\hbar}\operatorname{Im}\operatorname{Tr}
\!\left[h_{\mathbf r\mathbf r'}P_{\mathbf r'\mathbf r}\right],
\label{eq:local_chern_current}
\end{equation}
where $\hat{P}$ projects onto occupied states with $E<\mu=0$, using $V_0=10^5$, and $\hat J_i=-i[\hat X_i,\hat H_0]$ is Hermitian ($\hbar=1$), $A_c$ is the area of a unit cell, $q_i$ is the plotted site quadrature, and $j_{\mathbf r\to\mathbf r'}$ is the conserved current on an oriented bond.
The first quantity in Eq.~\eqref{eq:local_chern_current} is the real-space density whose spatial average gives the Chern number in a clean or sufficiently self-averaging bulk region.
For the quantitative diagnostic quoted in the main text, we average $C(\mathbf r)$ over a central window excluding the boundary region, and then over disorder realizations.
In these finite systems, the bulk-window average approaches the quantized values $\langle C(\mathbf r)\rangle_{\rm bulk}\to1$ for $W<W_c$ and $\to0$ for $W>W_c$ deep in each phase where $\xi\ll L$, and crosses $\langle C(\mathbf r)\rangle_{\rm bulk}\simeq0.5$ near the TM critical concentration where $\xi$ diverges.
Thus the color scale in Fig.~\ref{fig:random_percolation}(d)--(f) shows where the occupied projector locally retains the Chern-insulator character and where the marker changes sign or is suppressed by disorder. The plotted arrows show $q_i(\mathbf r)$, not the conserved bond current. The physical site-averaged current uses the real part of this matrix element; the plotted imaginary quadrature does not establish a conserved-current network. The local Chern marker fluctuates most strongly near the transition.

\subsection{Localization length calculation}

We compute the quasi-one-dimensional localization length $\Lambda$ by the TM method in a strip of width $N_y$ and length $N_x\gg N_y$~\cite{mackinnon1983scaling}.
The dimensionless quantity is $\Lambda/N_y$.
In a localized phase, $\Lambda/N_y$ decreases with increasing $N_y$; in an extended regime it increases; and at a scale-invariant critical point curves for different $N_y$ cross.
For the main-text data we use long strips with $N_x$ up to $10^6$ and estimate uncertainties from the variance accumulated in the TM iteration~\cite{wang2024anderson}.

Figure~\ref{fig:random_percolation}(b) shows the representative bimodal result and identifies $W_c\simeq 0.16$ for the parameters used in the main text.
The apparent broadened crossing for the smallest widths is a finite-size effect of the bimodal problem; it sharpens toward a single critical point as $N_y$ increases.
Because Eq.~\eqref{eq:bimodal_sigma_relation} is only a heuristic density-and-second-moment matching, we do not interpret this direct { bimodal} estimate as a controlled prediction of the positional-disorder SCBA.

\begin{figure}[tbp]
\centering
\includegraphics[width=1.0\columnwidth]{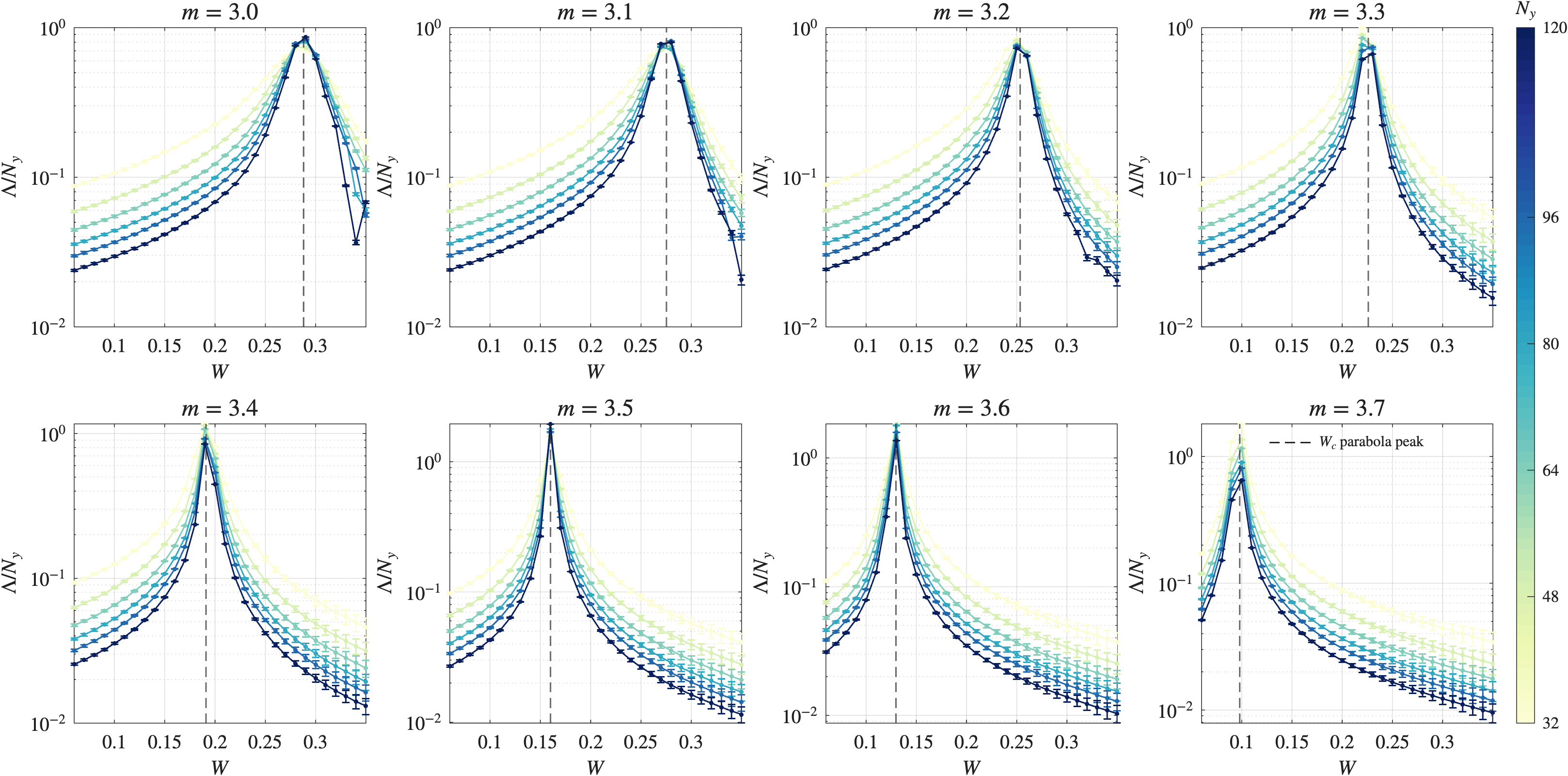}
\caption{Mass dependence of the normalized quasi-one-dimensional localization length for the bimodal vacancy problem.
Each panel shows $\Lambda/N_y$ as a function of vacancy concentration $W$ for a fixed clean mass $m$, with $m=3.0,3.1,\ldots,3.7$ arranged in a $2\times4$ grid.
Different curves correspond to strip widths $N_y=32,48,64,80,96,120$, and the dashed vertical line marks the estimated transition point extracted from the peak of a local parabolic interpolation of the large-$N_y$ data.
The TM length is $N_x=10^5$, the vacancy potential is $V_0=10^2$, and the disorder grid uses 30 equally spaced values of $W$ between $0.06$ and $0.35$.
At the representative mass $m=3.5$, $V_0=10^2$ is already in the vacancy-saturated regime: the two single-vacancy energies differ from their $V_0=10^5$ values by less than $6\times10^{-3}$.
The crossing/turning structure persists throughout this mass window, supporting the robustness of the vacancy-driven localization transition.}
\label{fig:supp_msweep_localization_length}
\end{figure}

Figure~\ref{fig:supp_msweep_localization_length} extends the same TM calculation to $m=3.0,3.1,\ldots,3.7$, which supports Figure~\ref{fig:random_percolation}(c) in the main text.
Each panel shows $\Lambda/N_y$ versus $W$ for $N_y=32,48,64,80,96,120$.
The dashed line marks the finite-size transition estimate from the large-width peak/crossing structure.
The persistence of this structure across the mass sweep shows that the vacancy-driven transition is not tied to a single fine-tuned value of the clean mass.

\begin{figure*}[tbp]
\centering
\includegraphics[width=1.0\textwidth]{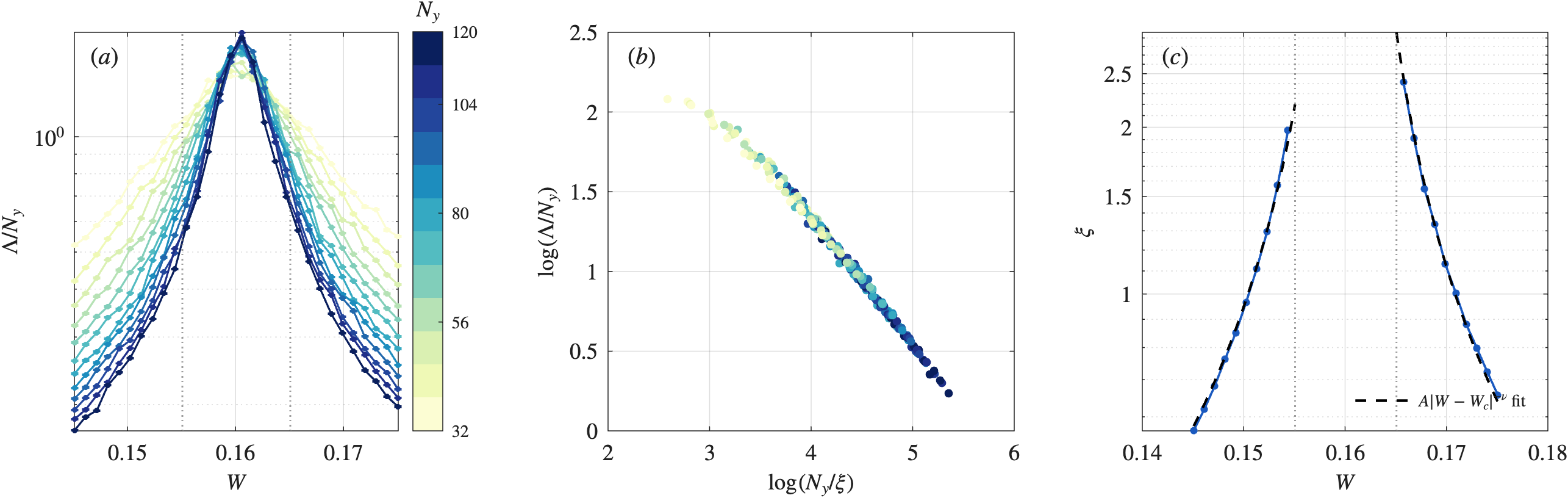}
\caption{Finite-size scaling analysis of the quasi-one-dimensional localization length at clean mass $m=3.5$.
The left panel shows the normalized localization length $\Lambda/N_y$ as a function of vacancy concentration $W$ in a narrow window around the crossing, using strip widths $N_y=32,40,\ldots,120$.
The middle panel shows the collapse of the same data after optimizing the correlation length $\xi(W)$, separating points on the two sides of the fitted critical disorder.
{ The right panel fits the optimized correlation length to $\xi=A|W-W_c|^{-\nu}$ outside the excluded central window, giving $W_c=0.161\pm0.002$ and $\nu=1.30\pm0.10$, including conservative fit-systematic uncertainties.}}
\label{fig:supp_m3p5_localization_xi_fit}
\end{figure*}

For a sharper finite-size scaling analysis, we use the MacKinnon-Kramer one-parameter scaling form~\cite{mackinnon1983scaling} \cite{MacKinnonKramer1981}
\begin{equation}
\frac{\Lambda(W,N_y)}{N_y}
= F_{\pm}\!\left(\frac{N_y}{\xi(W)}\right),
\qquad
\xi(W)=A_{\pm}|W-W_c|^{-\nu},
\label{eq:mk_scaling}
\end{equation}
with separate branches on the two sides of the transition.
In Fig.~\ref{fig:supp_m3p5_localization_xi_fit}, for $m=3.5$, we first locate the crossing window of $\Lambda/N_y$ versus $W$.
We then assign an optimized correlation length $\xi(W)$ to each disorder value so that all widths collapse onto the two universal branches $F_{\pm}$.
Finally, the optimized $\xi(W)$ is fitted to the power-law divergence in Eq.~\eqref{eq:mk_scaling}, excluding the narrow central region where finite-size saturation is strongest.
{ Allowing conservatively for variations of the fitting window, excluded near-critical points, and strip-width range, we quote $W_c=0.161\pm0.002$ and $\nu=1.3\pm0.1$ for this mass.}
 The TM and leading multifractal analyses both give accessible-size estimates $\nu\simeq1.3\pm0.1$. For the multifractal result, the uncertainty is a conservative allowance based on conditional fit errors and size-range variation, as detailed below. Neither analysis establishes the asymptotic universality class; a crossover toward the class-A value $\nu\approx2.6$~\cite{AltlandZirnbauer1997,SlevinOhtsuki2009} remains possible at larger sizes.

\subsection{Multifractal analysis}

 We use the local-extremum procedure of Ref.~\cite{moreno2023topological} within multifractal finite-size scaling~\cite{Rodriguez2011Multifractal}, retaining the explicit logarithmic normalization appropriate to the fixed-site observable defined below.
For a normalized eigenstate $\psi$ on an $L\times L$ sample, we compute
\begin{equation}
P_q=\sum_{\mathbf{r},\alpha} |\psi_{\mathbf{r}\alpha}|^{2q},
\qquad
\tilde{\tau}_q(L,W)=+\frac{\log \langle P_q\rangle}{\log L}.
\label{eq:tauq_definition}
\end{equation}
The average $\langle ... \rangle$ is over disorder samples and over the selected low-energy states.
For $q=1/2$, $\tilde{\tau}_q$ is larger for more spatially extended critical states, so a finite-size extremum of $\tilde{\tau}_q(W)$ provides a practical estimator of the critical disorder.

At each $L$ we fit the neighborhood of this extremum to a local cubic form,
\begin{equation}
\tilde{\tau}_q(L,W)
=\tilde{\tau}_q^c(L)
+\frac{1}{2}C_q(L)[W-W_*(L)]^2
+B_q(L)[W-W_*(L)]^3,
\label{eq:tauq_curvature_fit}
\end{equation}
where $W_*(L)$ is the finite-size critical point and $C_q(L)$ is the curvature.
The cubic term accounts for asymmetry of the finite-size curves and is not used as an independent critical exponent.
The localization-length exponent is then obtained from the curvature scaling
\begin{equation}
|C_q(L)|=\frac{c\,L^{2/\nu}}{\log L}\bigl(1+dL^{-y_\nu}\bigr),
\label{eq:tauq_curvature_scaling}
\end{equation}
 The leading fit sets $d=0$. For the fixed-site resolution used here, $\langle P_q\rangle=L^{-\tau_q}\mathcal F_q(x)$, with $x=(W-W_c)L^{1/\nu}$, gives
\[
C_q(L)=\frac{L^{2/\nu}}{\log L}
\left.\frac{d^2\log\mathcal F_q(x)}{dx^2}\right|_{x=x_*(L)},
\qquad x_*(L)=[W_*(L)-W_c]L^{1/\nu}.
\]
If the extremum approaches a fixed scaling coordinate with nonzero curvature, this yields Eq.~\eqref{eq:tauq_curvature_scaling}. The factor $2/\nu$ arises from two derivatives with respect to $W$.
 For $m=3.5$ and $q=1/2$, a weighted fit of the leading form of Eq.~\eqref{eq:tauq_curvature_scaling} to the curvatures at twelve system sizes gives $\nu=1.266$. We quote $\nu_{\rm MF}\simeq1.3\pm0.1$, with a conservative uncertainty accounting for fit errors and variation of the fitted size range.

\begin{figure*}[tbp]
\centering
\includegraphics[width=1.0\textwidth]{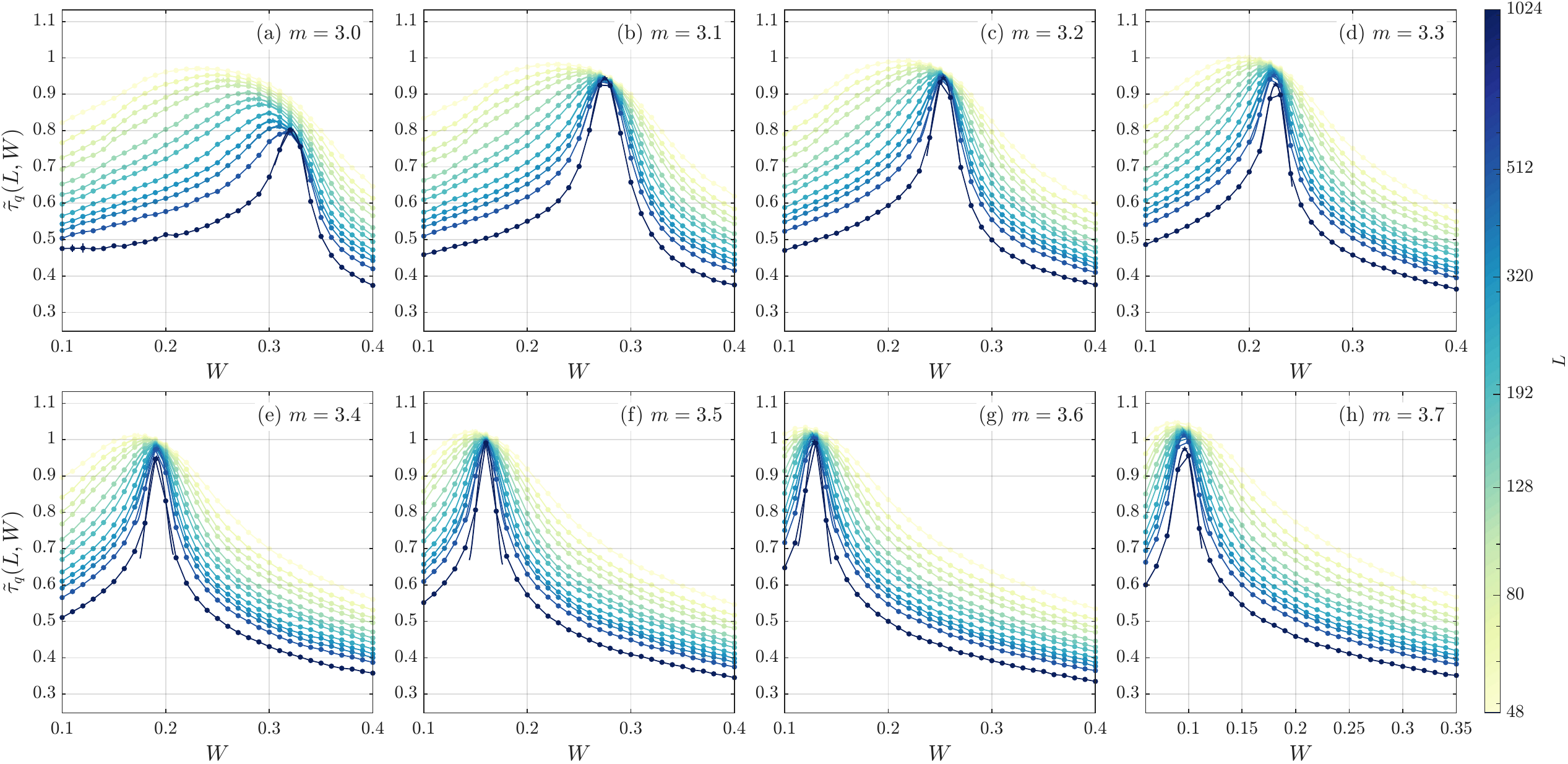}
\caption{Mass dependence of the multifractal finite-size curves near the vacancy-driven transition.
Each panel shows the disorder-averaged exponent $\tilde{\tau}_{q}(L,W)$ at $q=1/2$ as a function of vacancy concentration $W$ for a fixed clean mass $m$, with $m=3.0,3.1,\ldots,3.7$ arranged from left to right and top to bottom.
Different colors denote the linear sizes $L=48,64,80,96,128,160,192,256,320,384,512,1024$.
Markers with error bars show the aggregated disorder average and its standard error, while the smooth local fit segments and pentagons indicate the fitted finite-size extrema used to track the critical drift $W_{*}(L)$.
The systematic motion of these extrema with $m$ provides an independent multifractal diagnostic of the transition across the same mass sweep used in the localization-length analysis.}
\label{fig:supp_tauq_mass_sweep_grid}
\end{figure*}

\begin{figure}[tbp]
\centering
\includegraphics[width=0.95\textwidth]{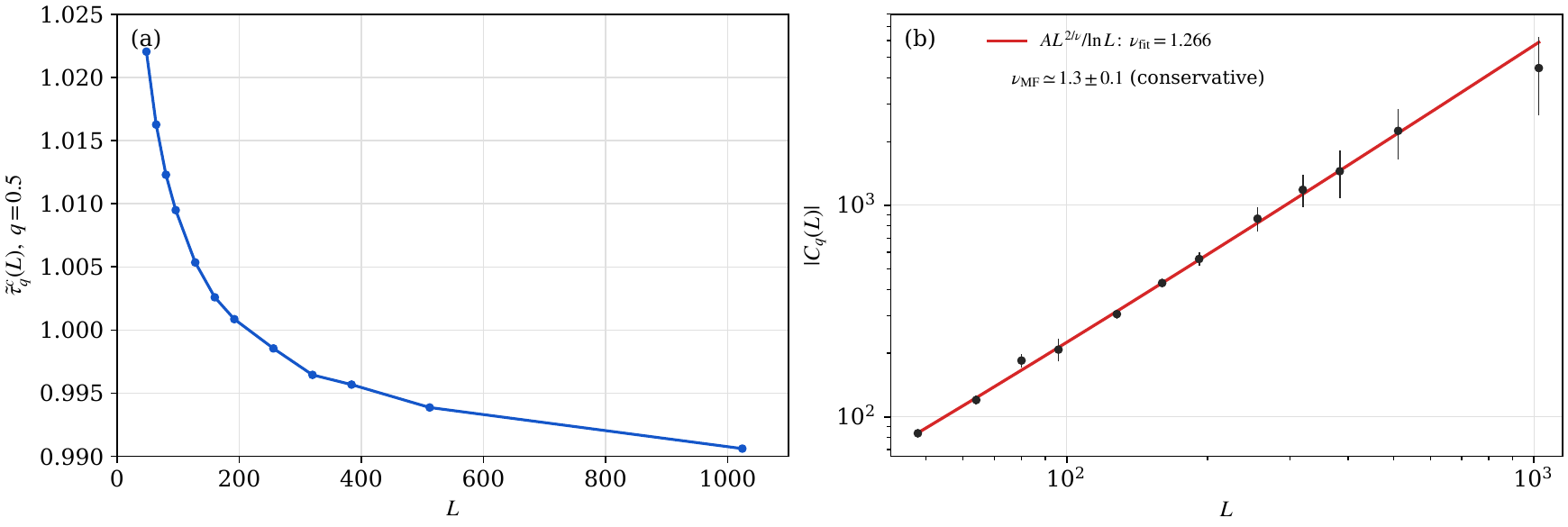}
\caption{Finite-size extraction of the critical exponent from the multifractal observable $\tilde{\tau}_{q}$ for $q=0.5$ at clean mass $m=3.5$.
 (a) Values of $\tilde{\tau}_{q}^{c}(L)$ at the fitted finite-size extrema. (b) Absolute curvatures $|C_q(L)|$ and the corrected leading fit $cL^{2/\nu}/\log L$. The full-data fit gives $\nu=1.266$, summarized as $\nu_{\rm MF}\simeq1.3\pm0.1$ with a conservative allowance based on conditional fit errors and size-range variation. Curvature error bars are retained from the existing extraction; no updated bootstrap uncertainty is implied. The attempted corrected fit including an irrelevant correction did not converge.}
\label{fig:supp_tauq_critical_exponents_m3p5}
\end{figure}

Figure~\ref{fig:supp_tauq_mass_sweep_grid} shows $\tilde{\tau}_{1/2}(L,W)$ for the same mass sweep as Fig.~\ref{fig:supp_msweep_localization_length}.
For $m=3.5$, the long-strip localization analysis gives $W_c^{\rm TM}=0.161\pm0.002$, while extrapolating the square-sample multifractal extrema gives $W_c^{\rm MF}\simeq0.161$.
The pentagons mark the fitted extrema $W_*(L)$.
Their systematic drift with $m$ gives an independent multifractal estimate of the transition line.
Figure~\ref{fig:supp_tauq_critical_exponents_m3p5} focuses on $m=3.5$.
The two panels show the finite-size critical values $\tilde{\tau}_{1/2}^c(L)$ and the fit of $|C_{1/2}(L)|$ using Eq.~\eqref{eq:tauq_curvature_scaling}.
 The corrected leading curvature fit is summarized by $\nu_{\rm MF}\simeq1.3\pm0.1$, with the conservative uncertainty allowance described above. Only the corrected leading fit is shown; a converged fit including the irrelevant correction has not been obtained.

\section{Critical exponent at nonzero energy and broken PHS}
\label{sup:finite_energy_exponent}

\begin{figure}[h!]
\centering
\includegraphics[width=0.98\textwidth]{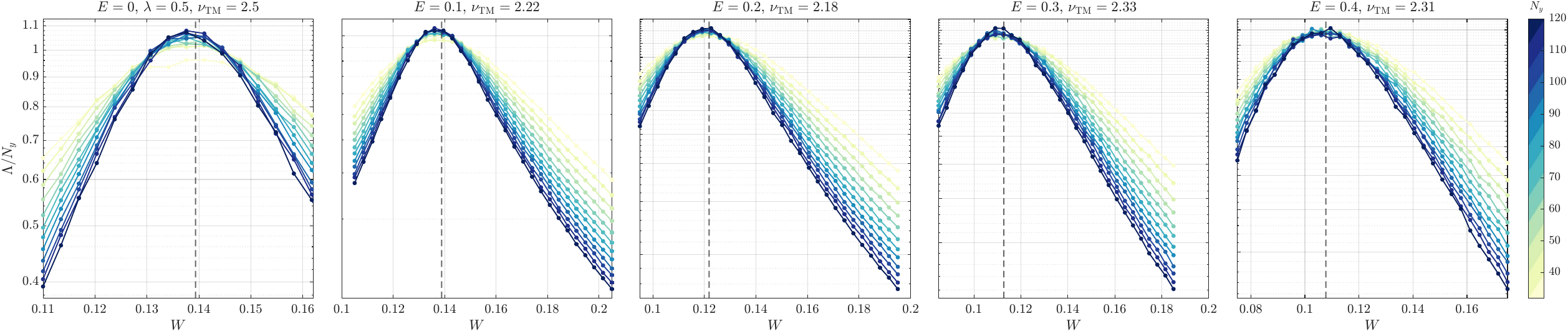}
\caption{TM localization scans for the vacancy-disordered QWZ model away from the undeformed zero-energy point.
The leftmost panel shows $E=0$ with the symmetry-breaking deformation $\lambda=0.5$; the remaining panels show $E=0.1$, $0.2$, $0.3$, and $0.4$ with $\lambda=0$.
Each panel plots the normalized quasi-one-dimensional localization length $\Lambda/N_y$ as a function of vacancy concentration $W$.
All panels use the twelve strip widths $N_y=32,40,\ldots,120$; the $E=0$, $\lambda=0.5$ panel is restricted to the analysis interval $0.11\leq W\leq0.162$, with the fluctuating large-$W$ tail omitted from the fit.
The longitudinal length is $N_x=10^6$ throughout.
The panel titles report the TM exponent obtained by fitting the optimized correlation length to $\xi=A|W-W_c|^{-\nu_{\rm TM}}$.
The black vertical dashed line in each panel marks the corresponding fitted $W_c$.
All panels use $m=3.5$, $t=2$, and $v=0.8$.
The vertical range is adjusted independently in each panel to display the full finite-size evolution compactly.}
\label{fig:supp_finite_energy_localization}
\end{figure}

The critical behavior at nonzero energy and under an explicit PHS-breaking deformation provides a direct test of the role of the special spectral symmetry at $E=0$.
Figure~\ref{fig:supp_finite_energy_localization} compares a scan at $E=0$ with $\lambda=0.5$ against scans at $E=0.1$, $0.2$, $0.3$, and $0.4$ with $\lambda=0$, all for the same clean mass $m=3.5$.
The size dependence of the normalized quasi-one-dimensional localization length moves systematically through the critical region in all five panels.

Separate fits give $\nu_{\rm TM}=2.22$, $2.18$, $2.33$, and $2.31$ at $E=0.1$, $0.2$, $0.3$, and $0.4$. Their spread is consistent with an accessible-size value near $2.3$; no joint fit or statistical combined uncertainty is implied.
For the symmetry-breaking deformation at $E=0$ and $\lambda=0.5$, we use all twelve strip widths $N_y=32,40,\ldots,120$, restrict the displayed analysis to $0.11\leq W\leq0.162$ to remove the fluctuating large-$W$ tail, and retain the resulting single connected collapse. The panel reports an effective crossing near $W_c\simeq0.139$ and $\nu_{\rm TM}=2.5\pm0.2$.
{ Both the finite-energy estimates and the $E=0$, $\lambda=0.5$ result are much closer to the integer-quantum-Hall value $\nu\simeq2.6$~\cite{SlevinOhtsuki2009} than the undeformed zero-energy estimate $\nu_{\rm TM}=1.3\pm0.1$.}
The shift is consistent with a symmetry-dependent finite-size crossover, but does not determine the asymptotic fixed point.

\end{document}